# Data Mining and Computational Screening of Rashba-Dresselhaus Splitting and Optoelectronic Properties in Two-Dimensional Perovskite Materials

Robert Stanton[1*], Wanyi Nie[2,3], Sergei Tretiak[3,4,5], and Dhara J. Trivedi[1*]

[1] *Department of Physics, Clarkson University, Potsdam, New York 13699, USA*

[2] *Department of Physics, University at Buffalo, Buffalo, New York 14260, USA*

[3] *Center for Integrated Nanotechnologies, Los Alamos National Laboratory, Los Alamos, NM 87545, USA*

[4] *Theoretical Division, Los Alamos National Laboratory, Los Alamos, NM 87545, USA*

[5] *Center for Nonlinear Studies, Los Alamos National Laboratory, Los Alamos, NM 87545, USA*

AUTHOR INFORMATION

* Corresponding authors: dtrivedi@clarkson.edu and robertmstanton@proton.me

# Abstract

Recent developments highlighting the promise of two-dimensional perovskites have vastly increased the compositional search space in the perovskite family. This presents a great opportunity for the realization of highly performant devices, and practical challenges associated with the identification of candidate materials. High-fidelity computational screening offers great value in this regard. In this study, we carry out a multiscale computational workflow, generating a dataset of two-dimensional perovskites in the Dion-Jacobson and Ruddlesden-Popper phases. Our dataset comprises ten B-site cations, four halogens, and over 20 organic cations across over 2,000 materials. We compute electronic properties, thermoelectric performance, and numerous geometric characteristics. Furthermore, we introduce a framework for the high-throughput computation of Rashba-Dresselhaus splitting. Finally, we use this dataset to train machine learning models for the accurate prediction of band gaps, candidate Rashba-Dresselhaus materials, and partial charges. The work presented herein can aid future investigations of two-dimensional perovskites with targeted applications in mind.

# Introduction

Halide perovskite materials are well-established in the literature for their applications in photovoltaic (PV) solar cells, light-emitting diodes (LEDs), radiation detection, and piezoelectric applications.[1–7] A natural emphasis has been placed on their usage in PV devices, owing largely to their low-cost building components and best-in-class power conversion efficiencies (PCEs) reaching as high as 25.7%.[8–11] Additionally, emerging uses of halide perovskite-based materials for spintronic, photonic, and ferroelectric applications highlight the continuing evolution of the field despite its maturity in the literature dating back to 2009.[12–18] One notable development is the synthesis of halide perovskites of reduced dimensionality, with two-dimensional perovskites (2DPKs) being the most extensively studied.[19–25] The 2DPKs have been demonstrated to confer increased structural stability to otherwise unstable perovskite systems through several mechanisms, such as out-of-plane distortions permitting otherwise problematic combinations of building components (as assessed by bulk metrics such as the octahedral factor $\mu$ and Goldschmidt tolerance factor $\tau_g$), and the incorporation of long spacing cations to alleviate hydrothermal degradation pathways.[26–29]

The two-dimensional analog of a bulk $ABX_3$ perovskite is formed by introducing organic spacing cations which separate the adjacent perovskite layers both electronically and spatially, forming a quantum well structure. The stacking arrangement and charge of the spacing cation dictates the resulting phase, with the Ruddlesden-Popper (RP) and Dion-Jacobson (DJ) phases being the most common throughout in the literature. For a more extensive review of 2DPKs for semiconductor applications, we refer the reader to Blancon *et. al.*[30] The DJ phase has a chemical formula of A'$A_{n-1}B_nX_{3n+1}$ with A' being the spacing organic cation of +2 formal charge, *n* being the inorganic layer thickness, A being a singly charged cation, B being a divalent cation, and X being a halogen.[31] The RP phase typically does not share A' cations between layers, resulting in a staggered stacking arrangement. This phase has a chemical formula of A'$_2$A$_{n-1}$B$_n$X$_{3n+1}$ with the A' cation carrying a formal charge of +1.[21] As noted by Marchenko *et. al*, exceptions to this phase distinction exist, however, they are very uncommon.[32] Increasing the structural diversity of 2DPKs with the freedom of choice in phase, layer thickness, and A' cation vastly expands the search space for highly performant materials when compared with their bulk counterparts. This highlights the need for computational investigations into these materials to allow for the coarse assessment of a broad range of material properties across a sufficiently representative sample of building components. Several investigations either manually parse the literature for 2DPKs or use freestanding monolayer analogs to analyze these materials in large quantities.[32–39] To our knowledge, however, no comprehensive dataset exists that includes fully represented 2DPKs constructed within a unified computational framework, assessing a broad range of properties from first principles. This study aims to fill this knowledge gap, enabling an accurate assessment of relative trends across the class of 2DPKs regarding their performance metrics for optoelectronic, spintronic, and thermoelectric applications.

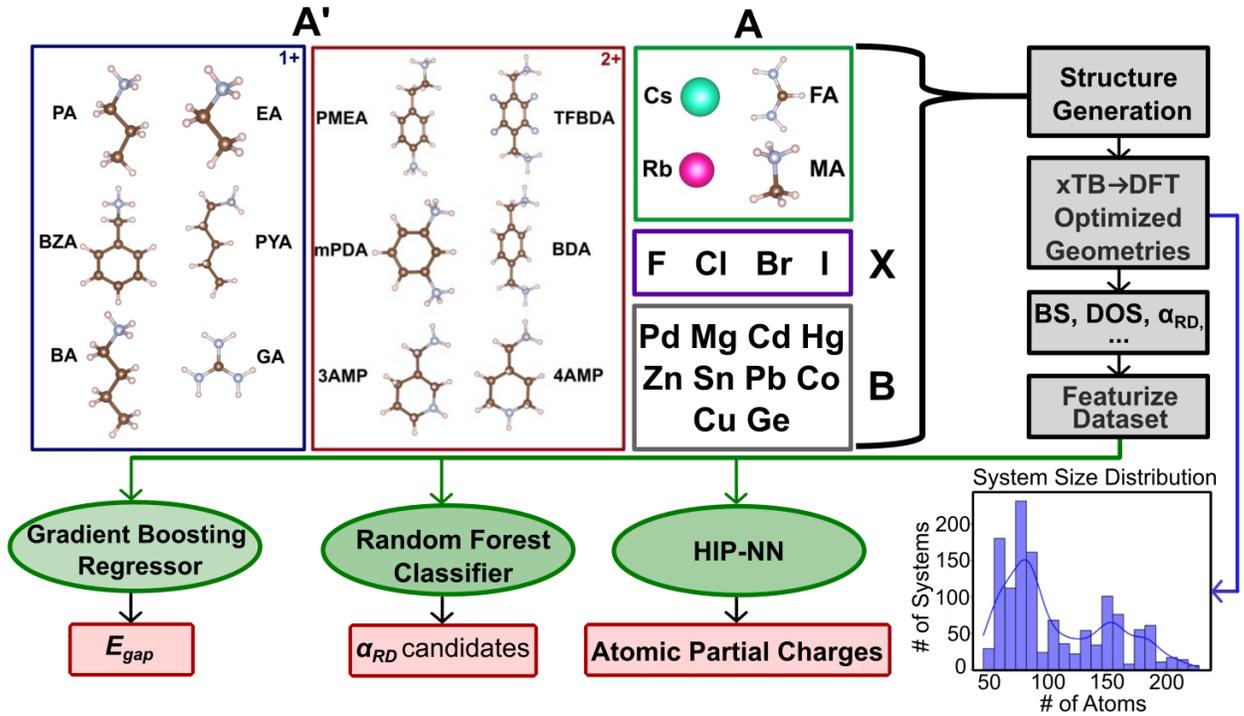

*Figure 1. Schematic overview of the project detailing the selection of building components, a rough outline of the computational workflow, as well as the resulting dataset and three ML models for prediction of electronic band gaps, Rashba-Dresselhaus splitting candidates, and atomic partial charges. The picture does not represent an exhaustive list of A', and A-site cations used in the present study, all building components can be found in the Dataset Design section.*

In this study, we employ a multiscale computational approach to investigate the structural, optoelectronic, and thermoelectric properties of 2DPKs in the DJ and RP phases. The key focus of the work is to present the first large-scale 2DPK database in which all structures and computed quantities are obtained through a unified computational farmwork. Additionally, the RashbaPy code presented herein serves to facilitate future investigations into spintronic applications across a broad range of materials including but not limited to 2DPKs. Finally, the machine learning models for electronic band gaps, Rashba-Dresselhaus splitting, and atomic partial charges can be used to aid in future computational investigations to drastically accelerate materials discovery. The work presented here can be condensed to three main points, which are summarized schematically in Fig. 1:

1. We combine density functional theory (DFT) calculations together with additional frameworks which leverage the *ab initio* data, such as the Boltzmann Transport Equations (BTEs) to assess material characteristics relevant to diverse and emerging application scope of 2DPKs. An exhaustive list of the computed quantities is available in Supplementary Tables S2 and S6.

2. We implement a high-throughput computational framework for the assessment of Rashba-Dresselhaus (RD) splitting from *ab initio* calculations, RashbaPy, which is an open-source Python package available at https://github.com/r2stanton/rashbapy. We then use this to compute RD splitting for a large subset of the perovskites in the present study.

3. Using this newly presented dataset, we generate machine learning models for the prediction of electronic band gaps ($E_{gap}$), Rashba-Dresselhaus splitting ($\alpha_{RD}$), and atomic partial charges, suitable for usage across a broad range of the periodic table. In conjunction with these models, we compute the Pearson correlation coefficient (PCC) between all descriptors used in the ML models (Figure S12). The PCC serves as a metric for quantifying the linear correlation between two descriptors, with +1.00 (-1.00) denoting a perfectly linear and positive (negative) correlation.

## Dataset Design

In generating the dataset, we employ the eight most common A-site cations, comprising both inorganic elements and organic molecules: cesium (Cs), rubidium (Rb), methylammonium (MA), formamidinium (FA), azetidinium (AZ), methylhydrazinium (MHy) and dimethylammonium (DMA). The utilized B-site cations include several selections which are well-established in the halide perovskite literature such as lead (Pb), tin (Sn), and germanium (Ge), as well as less well-explored elements like palladium (Pd), magnesium (Mg), cobalt (Co), cadmium (Cd), mercury (Hg), zinc (Zn) and copper (Cu). We consider the four standard halogens used in perovskite systems of fluorine (Fl), chlorine (Cl), bromine (Br) and iodine (I). Furthermore, addressing the structural diversity offered to 2DPKs over their bulk counterparts, we investigate 14 spacers in total with seven specifically used in the DJ phase: 3-(Aminomethyl)piperidinium (3AMP) 4-(Aminomethyl)piperidinium (4AMP), 2-(4-aminophenyl)ethylamine (PMEA), thieno[3,2-b]thiophene-2,5-diyldimethanammonium (TTDMA), tetrafluorobenzenedimethanammonium (TDFBDA). Another seven spacers are typically used in the RP phase: butylammonium (BA), propylammonium (PA), guanidinium (GA), 2-amino-methylthiophene (ThMA), 2-thiopheneformamidine (ThFA), benzylammonium (BZA), and p-fluorobenzy-lammonium (pFBA). For the DJ and RP phases we consider layer thicknesses $n$=1-5 and $n$=1,2, respectively. The latter systems typically have large unit cells compared to that of the DJ phase and their modeling is thus limited by numerical expense. The median system size of the 2104 converged DFT geometries presented in this work is 88 atoms, with the smallest (largest) system consisting of 46 (228) atoms (Fig. 1). Full details regarding computed quantities, refinement of the presented dataset, and the subsets used in all machine learning models are available in the Supplementary Information.

## Results and Discussion

### Structural Characteristics

Structural distortions in 2DPK systems are responsible for modulating several key characteristics associated with optoelectronic and spintronic applications within this class of materials.[40–42] Structural distortions in pristine perovskite systems (that is, those geometric features excluding effects of defects and dopants) are broadly characterized as either octahedral distortions, or structure-wide distortions such as ion dislocations and inter-octahedral tilting patterns.[43,44] Octahedral distortions have been demonstrated to be strong predictors of properties like the optoelectronic band gap, RD splitting, and ferroelectricity in 2DPK and bulk perovskite systems.[42,45–48] Figure 2(a-b, d-e) depicts the distribution of B-X bond lengths and *trans* X-B-X bond angles for all gapped perovskite systems in the present study as a function of B-cation and X-halogen. The strong correspondence between Pb- and Sn-containing perovskite distributions in Fig. 2(a,b) aids in explaining the highly similar optoelectronic properties they exhibit throughout the literature. Supplementary Figure S5 shows additional structural and octahedral distortions for the gapped 2DPK systems generated in the present investigation. Also of note is the prominence of large structural distortions in the F-containing perovskites as evidenced primarily by the large deviations from 180° in the *trans* X-B-X bond angles (Fig. 2d). This is explained by the small ionic radius of the fluoride ion resulting in octahedral tolerance factors, $\mu = \frac{R_B}{R_X}$, larger than those typically considered tolerable for bulk perovskites of $0.414 < \mu < 0.732$ (Supplementary Fig. S6).[49] Goldschmidt tolerance factors, $\tau_g = \frac{R_A+R_X}{\sqrt{2}(R_B+R_X)}$, for bulk perovskites are expected to be within $0.825 < \tau_g < 1.059$.[49] In these formulas, $R_A, R_B$ and $R_X$ are taken to be the ionic radii of A-, B-, and X-site ions respectively. The computations of all structural distortions and tolerance factors were carried out with the freely available Pyrovskite package. Our investigation here comports with previous studies suggesting that $\tau_g$ constraints are largely relaxed in the family of 2DPKs.[50,51] This is because, in 2DPKs, out-of-plane corrugation can alleviate strain caused by mismatched ionic radii, particularly for 2DPKs with low layer thickness *n*. This is demonstrated clearly by the Mg-containing perovskites, whereby numerous 2DPKs with $\tau_g > 1.1$ still result in octahedral distortions less than or similar to those associated with well-established Sn-, Pb-, and Ge-containing 2DPK systems (Fig. 2c and Supplementary Figs. S4 and S5). In contrast, our results demonstrate that 2DPKs do not relax constraints associated with the octahedral factor $\mu$ as effectively. This can be seen whereby the fluoride-containing 2DPKs with $\mu > 0.8$ (Supplementary Fig. S6) exhibit larger octahedral distortions across all metrics when contrasted with the other halogens in Fig. 2(d,e) and Supplementary Figs. S4 and S5. The subject of tolerance factors as applied to 2DPKs and perovskites of other phases is still under active development, with a number of candidate metrics having been introduced in addition to modifications of the preexisting Goldschmidt and octahedral tolerance factors.[49,50,52–54]

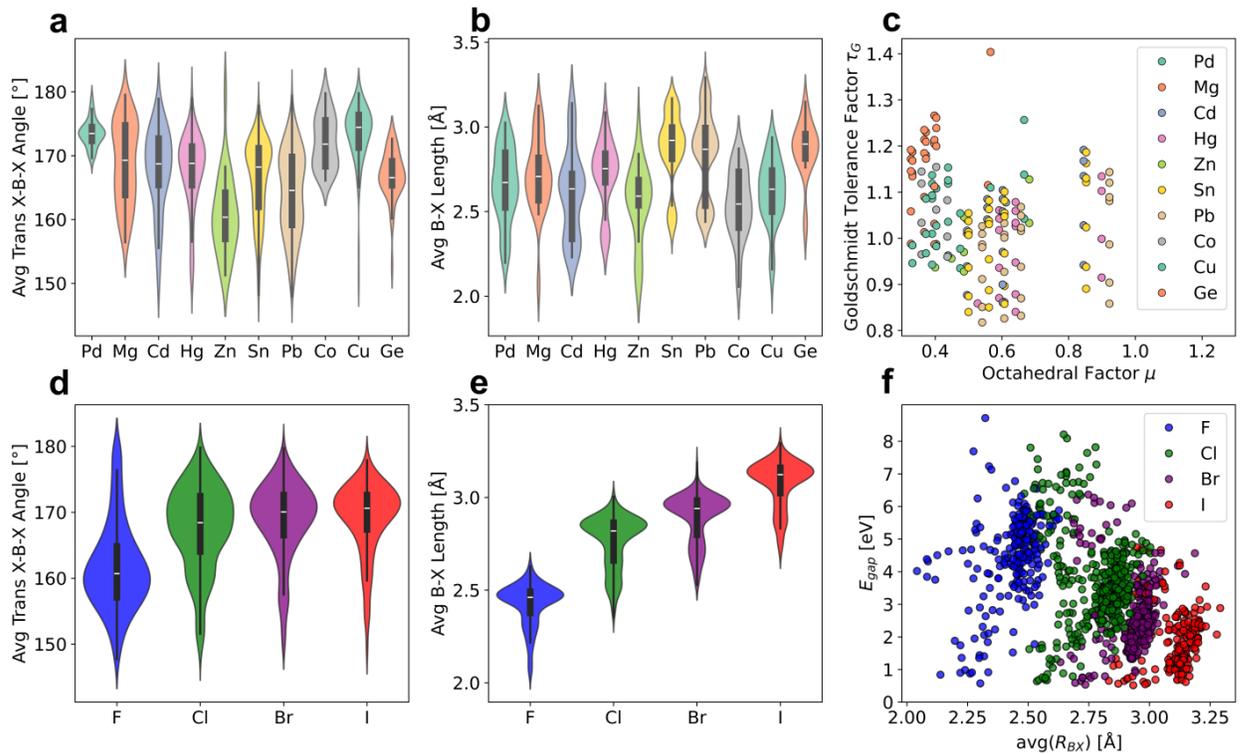

*Figure 2. Structural properties associated with the gapped perovskite systems in the newly generated dataset. (a, b) Distributions of trans B-X-B bonds and B-X bond lengths as a function of the B-site cation. Trans X-B-X bond angles in idealized octahedra correspond to 180°, (c) Scatter plot of the octahedral factor ($\mu$) vs. Goldschmidt tolerance factor ($\tau_G$) for the perovskites presented in the dataset. (d, e) Distributions of trans X-B-X bond angles and B-X bond lengths as a function of the X-site halogen. (f) Electronic band gaps as a function of the average B-X bond distance.*

Another structure-property relationship of note is the interplay between electronic band gaps and B-X bond lengths. The large peak in the B-X bond length distributions in Fig. 2e are attributed almost entirely to the Group 14 elements, which constitute a large portion of the dataset and exhibit strikingly similar geometric properties. Similarly the long tails of the distribution arise from the inclusion of less common B-site ions such as, Mg, Cd, Hg, Zn, Co, and Cu. As is well-established, electronic band gaps decrease with the increasing radius (and thereby increased B-X bond lengths) of the halogen as a general trend across both 2DPKs and bulk perovskites.[55] Figure 2f, however, indicates that within a halogen group increased B-X bond distances result in lower band gaps. The explanation for this is that, given a particular electron configuration, smaller bond distances lead to increasing overlap of electronic states between the B- and X-site ions. This increased overlap, in turn, facilitates electronic delocalization and enhances band dispersion, thereby reducing band gaps. A similar argument regarding band dispersion serves to explain the decreasing effective masses (as effective masses are inversely proportional to the curvature at the band edge) with increased ionic radii observed by Feng and Xiao in Cl-, Br-, and I-containing perovskites.[56] Establishing trends in structure-property relationships of this nature may aid in fine tuning materials selection for targeted applications. For example, the abovementioned interplay between halogen ionic radius, B-X bond lengths, and band dispersion allows for the selection of 2DPKs with tailored band gaps and low charge carrier effective masses for hybrid photovoltaic and thermoelectric applications.

## Electronic Properties

Structural deviations associated with octahedral distortions and combinations of different B- and X-site ions can give way to drastically modified optoelectronic properties with only modest composition changes to the 2DPK.[57,58] Accurate assessment of electronic properties, such as band gaps, in perovskite systems with diverse B- and X-site ions using *ab initio* methods requires careful consideration. In the case of Pb-containing perovskites, there is a natural error cancellation associated with the usage of typical GGA functionals like PBE and the neglect of spin-orbit coupling (SOC) noted in the literature.[59,60] This stems from the fact that SOC in Pb-containing perovskite systems is often exceptionally large, reducing the band gap by around 1 eV and leading to fortuitous agreement between GGA-DFT calculations and experimental band gaps. However, in perovskites which do not contain Pb as the B-site cation, band gap underestimation still occurs, as the spurious error cancellation is less reliable. With this being the case, all electronic band gaps in the present study are computed as follows:

$$E_{gap} = \widetilde{E_{GLLB-sc}} - \Delta_{SOC} \quad (1)$$

Here the $\widetilde{E_{GLLB-sc}}$ term corrects for a systematic underestimation of electronic band gaps associated with the PBE functional, and the $\Delta_{SOC}$ term accounts for the band splitting due to SOC, calculated as the difference between the PBE+SOC and PBE band gaps.[61] Full details regarding the GLLB-sc functional, DFT simulation parameters, and the SOC calculations can be found in the Supplementary Information. We find that the surrogate $E_{gap}$ estimates yield band gaps that are closer to the corresponding experimental values than the PBE gap, the PBE+SOC gap, or the GLLB-sc gap alone in most scenarios. An extensive list of comparisons between computed and experimental band gaps can be found in Supplementary Table S1. The observed performance is most accurate for Group 14 elements, with predicted gaps typically within 0.5 eV of the reference values. Computed gaps in perovskites containing less common B-site ions such as Hg, Zn, Cd, and Pd are, on average, less accurate compared to experimental numbers but still mostly within 1 eV of the reference value. Future studies aiming for a more fine-grained investigation of perovskites containing the latter elements would benefit from frameworks offering an on-site Coulomb correction, such as DFT+U to effectively treat the strongly correlated d-electrons that contribute to the band edges in these systems. Band gap distributions by B-site cation and X-site halogens are pictured in Fig. 3(a,b). This newly presented dataset contains 130 2DPK systems with band gaps within 0.4 eV of the ideal band gap of 1.34 eV for solar energy harvesting applications, represented by the shaded region in Fig. 3(a,b). The idealized 1.34 eV band gap is derived from the Shockley-Queisser model, which predicts the maximum PCE of single junction solar cells. We note that these 130 2DPKS are not restricted to typical Pb- or Sn-based 2DPKs, rather they span all four halogens and six B-site cations (Pb, Sn, Ge, Hg, Cd, and Pd). These 2DPKs are notable candidates for photovoltaic applications, which can aid future studies looking to employ lead-free perovskites with favorable optoelectronic properties. Thus, our results underscore the importance of further investigations into non-standard B-site cations, as many of them may prove promising for PV applications. Of the gapped systems in the present dataset, the remaining 1170 samples fall out of the ideal range for direct application as single junction photovoltaics. However, many of these 2DPKs still have various merits and possible areas of application including but not limited to LEDs, radiation detection, or the formation of mixed 3D-2D perovskite systems for improved hydrothermal stability.

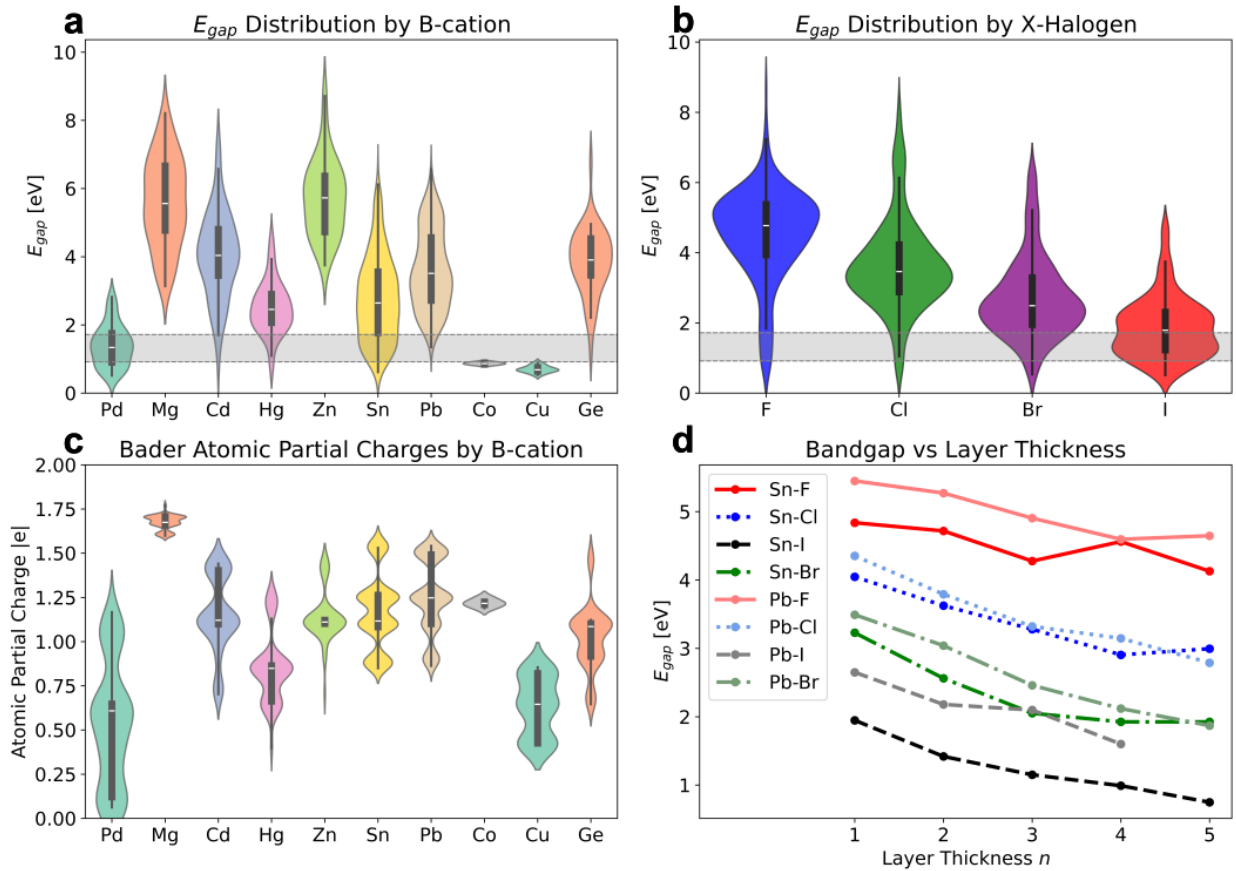

*Figure 3. Electronic properties associated with the gapped perovskite systems in the newly generated dataset. (a, b) $E_{gap}$ as a function of B-site and X-site ions for the gapped 2DPKs in the generated dataset respectively, (c) Atomic partial charges of the B-site cations across the dataset, and (d) $E_{gap}$ as a function of layer thickness for the Pb- and Sn-containing 2DPKs.*

Further investigation into the interplay between chemical and structural properties and the resulting optoelectronic characteristics reveals that the divalent oxidation state does not serve as a strong predictor of partial atomic charges in the resulting 2DPK system. Figure 3c illustrates the strongly heterogenous charge distributions across the B-site cations, despite their correspondence in oxidation state. Analogous to similarities in structural distortions previously discussed, the electronic structure of Pb- and Sn-containing perovskites are nearly identical, exhibiting remarkably similar distributions of atomic partial charge and electronic band gaps. Worth noting is that similarities in partial charge distributions do not always correlate with similar electronic band gaps distributions, as can be seen by using Co and Mg as an illustrative example. Both Co- and Mg- containing perovskites admit very narrow distributions of atomic partial charges, but widely different behavior in terms of their electronic band gap distributions. This counterintuitive notion is explained through the analysis of their corresponding geometric properties. Atomic partial charges will be determined largely by the local chemical environment of a given atom, for which the B-X bond length is the most relevant quantity. In other words, the partial charge of an atom should not be largely affected by minute structural variations such as small conformal changes to the octahedral structure. Figure S4 shows strikingly similar B-X bond length distributions of Co, and Mg which correlate with their narrowly distributed atomic partial charges. Electronic band gaps, on the other hand, are much more sensitive to minor distortions within the structure, as these changes can drastically impact the overlap of electronic states, and thereby the band dispersion. Figure S5 shows that conformal changes to the octahedra (indicated by $\Sigma$, and $\Lambda_2$) better explain the differences in band gap distributions between Co and Mg. The very narrow (broad) distribution of these two octahedral distortion parameters in Co (Mg) correlate directly with the narrow (broad) band gap distribution. This means that particular care must be given in interpreting trends across 2DPK systems by considering the sensitivity of the structure-property relationship under question. The Pd-containing 2DPK systems provide further support of this line of reasoning, whereby the broad distribution of atomic partial charges (Fig. 3c) has a minimal impact on the narrowly distributed electronic band gaps (Fig. 3a) because of the tightly clustered octahedral distortion parameters $\Sigma$ and $\Lambda_2$ in Fig. S5. Despite these exceptions, the charge-band gap heuristic remains statistically useful, as Fig. S7 shows a strong positive (negative) correlation between the average charge at the B-site (X-site) and the band gap, with a PCC of 0.73 (-0.75) for the gapped systems in the 2DPK dataset. In addition to this, our results corroborate the notion that as layer thickness in the perovskite layers increases, a downward trend is observed in the resulting band gap, approaching that of the bulk 2DPK analog (Fig. 3d). We found little to no relationship between geometric or chemical characteristics of the spacer molecule and the resulting band gap. For example, the spacer length and average Pauling electronegativity have little to no correlation with $E_{gap}$, with PCCs of 0.02 and -0.07 respectively. In other words, the spacer and its unique interactions with the adjacent perovskite layers are key factors in determining its effect on the resulting electronic structure (Supplementary Fig. S8). This underscores the necessity of developing complex descriptors to effectively characterize diverse interactions, such as inter-spacer π-π interactions, and non-covalent interactions between chemically active groups on the spacer and the adjacent perovskite layers. This is particularly difficult because the insights gained from a manual inspection of these quantities often relies heavily on chemical intuition which is difficult to formulate in an algorithmic and quantitative manner.[19,62–66] We emphasize that this data does not imply that the spacer molecules have little impact on the electronic structure of the perovskite systems. Instead, it suggests that a fine-grained inspection of individual systems is necessary to elucidate the complexities associated with these impacts.

## Thermoelectric Performance

The thermoelectric (TE) performance of 2DPKs is still an area of applicability under active investigation, which has demonstrated much promise in recent years. Given the inherent inefficiencies of PV solar cells, there is a growing need to explore other means to drive up their performance. The potential to harvest residual thermal energy makes the integration of TE materials into PV devices an increasingly important area of focus. To this end, using BoltzTraP2 (a solver for the BTEs in the relaxation time approximation) in combination with *ab initio* DFT data, we obtain thermoelectric power factors ($TPF$s) for the 2DPKs in the present dataset.[67] BoltzTraP2 utilizes eigenvalues from a uniform, densely sampled ground state *ab* initio calculation order to approximate Onsager coefficients (electronic conductivity, the electronic contribution to the thermal conductivity, and the Seebeck coefficient in the rigid band and relaxation time approximations. Further computational details of all BoltzTraP2 calculations are available in the Supplementary Information. The thermoelectric power factor is defined as $TPF = S^2\sigma$ where $S$ is the Seebeck coefficient, and $\sigma$ is the electrical conductivity of holes for *p*-doped configurations, and electrons for *n*-doped configurations. This metric assesses a semiconductor's ability to maintain temperature gradients and efficiently transport residual thermal energy coming from, for example, incident light that does not directly lead to the separation and extraction of excited charge carriers. This is particularly important for perovskites, where non-radiative losses drive a large portion of PV inefficiency. With this being the case, the search for hybrid PV-TE devices has increased in recent years, with 2DPKs serving as a particularly promising avenue of this investigation.

The $TPF$ is largely dependent on the severity of dispersion near band edges of a 2DPK given system, as this corresponds to large group velocities which are key quantities throughout the BTEs. This naturally leads to strong performance in cases with highly dispersive p-bands at the band edge demonstrated in Fig. 4a. In contrast, the Group 14 elements (whose conduction bands are dominated by Ge-4p, Sn-5p, Pb-6p electrons) produce the most efficient TE materials in *n*-doped configurations. Likewise, the relatively flat d-bands which constitute the conduction bands of the less commonly used B-site cations like Pd, Co, and Cu lead to poor $TPF$s (Fig. 4a). Typically, halogen p-bands dominate the valence band which directly results in improved $TPF$s of holes as the principal quantum number of the valence p-electrons (and thereby their band dispersion) increases uniformly across the halogens from fluorine to iodine (Fig. 4b). This explains why the most widely used 2DPK systems exhibiting p-p transitions across the band edge (Pb-I, Pb-Br, Sn-I, and Sn-Br) show the most promise for mixed PV-TE device implementations. Furthermore, it suggests that there is very little room for improvement in terms of TE performance with typical divalent cations. Additional improvements may require the consideration of more novel materials, such as lanthanide- or actinide-based perovskites.

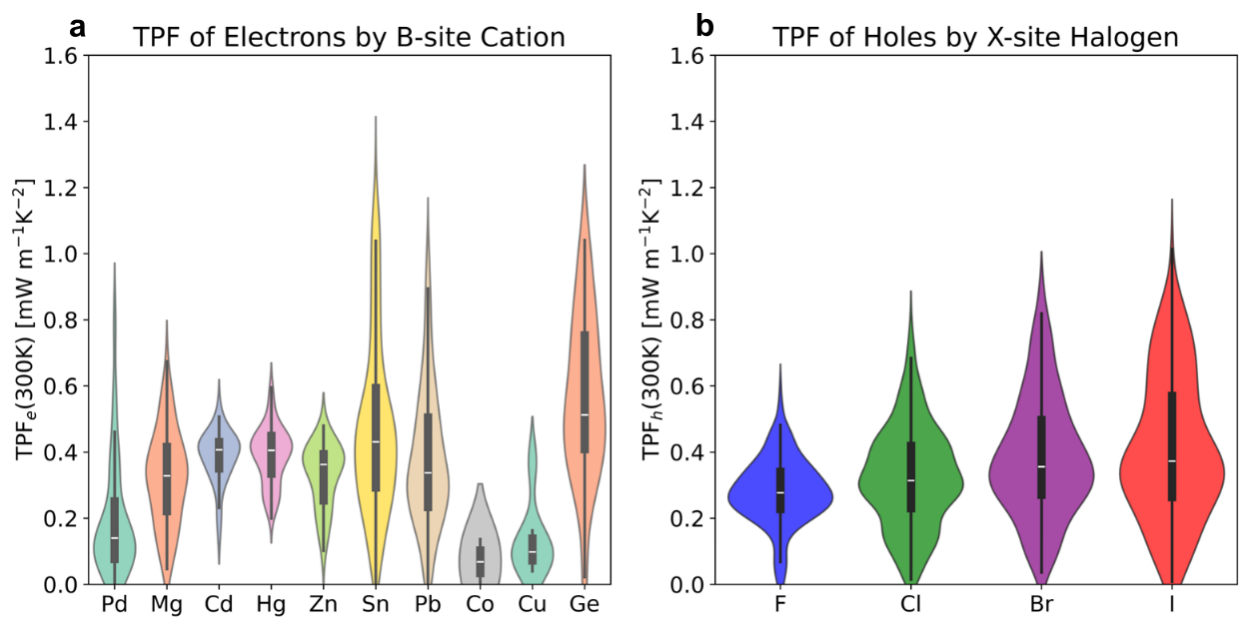

*Figure 4. Thermoelectric power factor of (a) electrons across all B-site cations, and (b) holes across all X-site halogens.*

The role played by the spacer in determining TE performance can largely be classified in two ways. First, as has been noted throughout the literature, band alignment between the eigenstates localized to the perovskite and spacer subsystems can lead to mid-gap electronic states which trap charge carriers on the spacer.[68–70] In these scenarios, spacer molecules can significantly suppress TE transport, as observed with the ThFA spacer, which introduces a trap state localized to the spacer just below the conduction band edge of the perovskite (Supplementary Fig. S9). The other means by which the spacer can modulate TE performance is by conferring additional stability to the adjacent perovskite octahedra. $TPF$s for both holes and electrons are negatively correlated with all octahedral distortion parameters, suggesting that perovskites, which exhibit near ideal symmetry, offer better TE performance.

## High-throughput simulation of Rashba-Dresselhaus Splitting

Another enticing property that stands out in the realm of 2DPKs is the possibility for their realization of RD splitting. Rashba and Dresselhaus splitting are two forms of band splitting thought to be induced by SOC in the presence of an inversion symmetry breaking event coupled with local or external electric fields. Several excellent reviews in the literature explain these forms of spin splitting at length. Rashba and Dresselhaus splitting are frequently referred to as two- and three-dimensional phenomena, respectively. In the context of 2DPKs, however, it is common for a hybrid form of RD splitting to occur, as has been observed by Maurer *et al.* as well as Bhattacharya and Kanai.[42,71] Figure 5a demonstrates a schematic and representative dispersion relation that occurs as a result of RD splitting. The offset of the band edge in reciprocal space from the nearest high symmetry point ($dk$), and the energy offset ($dE$) between the band edge and energy at the vertex from which splitting occurs are most common ways to obtain the RD splitting, $\alpha_{RD} = 2\frac{dE}{dk}$.[22,72] A common issue with computing RD splitting from *ab initio* methods is that very dense k-point sampling is needed to ensure that the shifted band edge is sufficiently close to one of the points used in the DFT calculation. To this end, we implemented quadratic interpolation of the bands ensuring that even modest k-point densities give reliable values for $dE$, and more importantly $dk$. Our code is freely available and has been benchmarked against RD splitting found in the literature (Supplementary Fig. S10 and Table S3).[72] Full details regarding the high-throughput computation of $\alpha_{RD}$ for the selected perovskites in the present study, as well as additional details regarding the RashbaPy code can be found in Supplementary Information. Given the ongoing debate regarding the origin, mechanism, and characterization of Rashba and Dresselhaus splitting in 2D and 3D perovskite systems, we discuss at length how we aimed to mitigate falsely identifying high $\alpha_{RD}$ candidate materials through careful manual inspection of all 2DPKs with $\alpha_{RD} > 1.0$ eV Å.

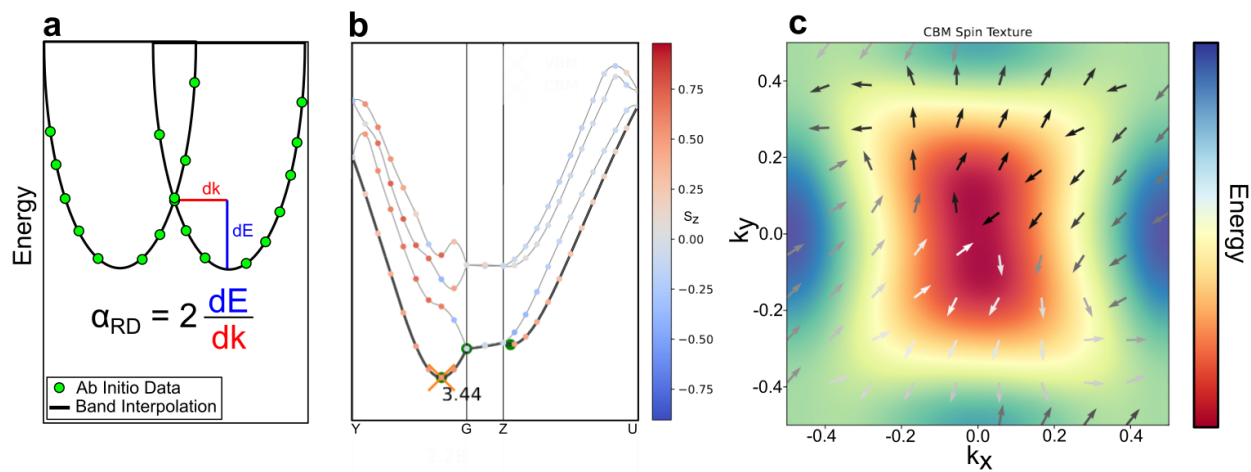

Figure 5. (a) A schematic representation of the algorithm used in the RashbaPy code for computing Rashba-Dresselhaus splitting ($\alpha_{RD}$) by way of interpolating ab initio band structure data. (b) An example 2DPK ($BDA_1MA_3Pb_4I_{13}$) from the dataset identified as exhibiting a large $\alpha_{RD}$ of 3.44 eV Å. (c) Spin texture generated from the RashbaPy code for manual inspection or further delineation of Rashba vs. Dresselhaus splitting.

Of the structures presented in the dataset, we find 45 which suggest the possibility for $\alpha_{RD} > 2.0$ eV Å, and five with $\alpha_{RD} > 3.0$ eV Å. As evidenced by the literature on RD splitting, the mechanisms by which the effect is induced remain highly elusive. The distribution of $\alpha_{RD}$ in Fig. 6 reaffirms the fact that heavy B-site ions are a necessary condition for R-D splitting. Naturally, the atomic number of the B-site cation has a large impact on the prevalence of $\alpha_{RD}$ because the SOC-induced splitting is result of the relativistic effects associated primarily with heavy atoms. Likewise, from Fig. 6 one can see that the organic spacers which consist entirely of relatively light atoms (up to sulfur), have little to no effect on the appearance of $\alpha_{RD}$ splitting in the material. We find no computed structural or chemical characteristics which exhibit a PCC with $\alpha_{RD}$ larger than 0.5 within the full dataset of 2DPKs. Similarly, when restricting analysis to single B-site cations, only weak positive correlations (PCC < 0.2) are found to be statistically significant regarding structural distortion parameters. One metric that stands out for predicting $\alpha_{RD}$ is the difference in Pauling electronegativity between the X-site and B-site ions (PCC of 0.42). This electronegativity difference, $X_{BX}$ Diff, is computed as $X_{BX}$ Diff $= X_B - X_X$, and is always negative for the 2DPKs in this dataset. $X_B$ and $X_X$ denote the Pauling electronegativity at the B- and X-site respectively. An interesting conclusion from this positive correlation is that compounds with weaker ionic interaction between the B- and X-site ions exhibit larger $\alpha_{RD}$. This, at first glance, seems counter-intuitive because ionically bonded compounds have a higher propensity towards polar configurations, and thereby local electric fields to drive up RD splitting. However, our results suggest it may be the case that the structural rigidity associated with these increasingly ionic B-X bonds may mitigate these polar distortions to begin with. In other words, 2DPKs which have prominent covalent character in the B-X bonds appear more likely to allow for distortions that correlate with $\alpha_{RD}$ than those with strong ionic B-X bonds.

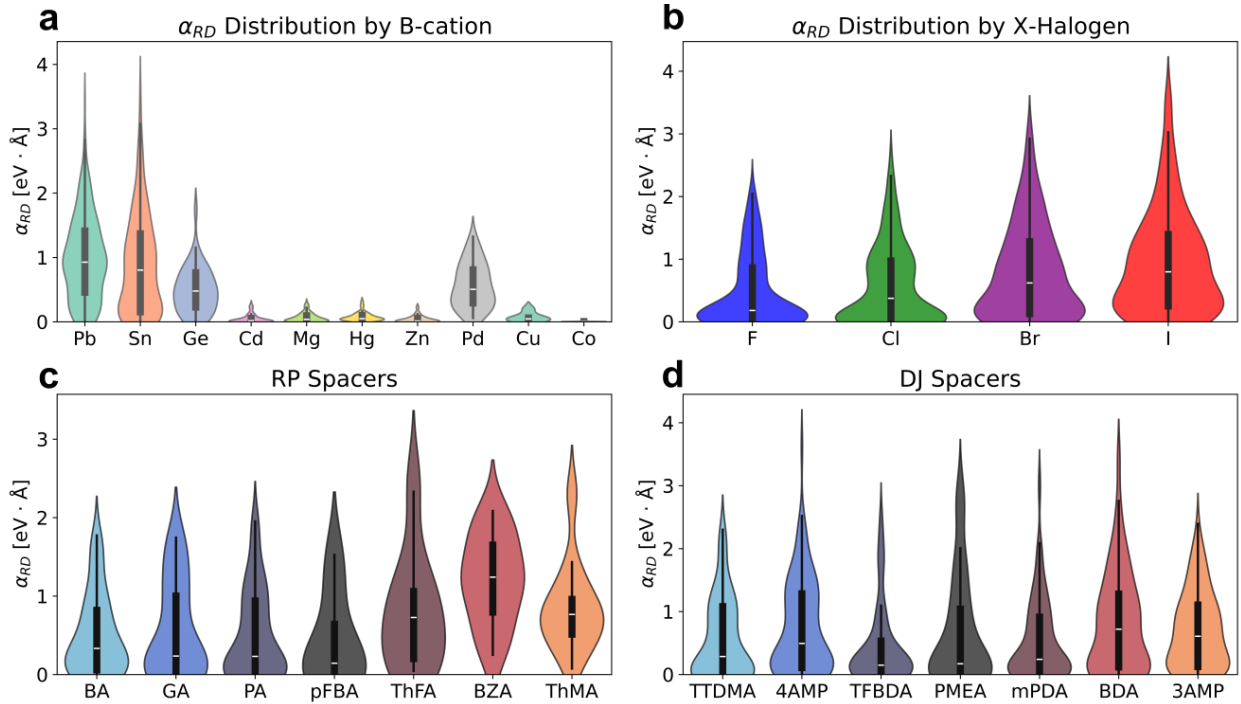

*Figure 6. Rashba-Dresselhaus splitting, $\alpha_{RD}$ by (a) b-site cation, (b) X-site halogen, (c) organic spacers used in the RP phase, and (d) organic spacers used in the DJ phase.*

The difficulty in attributing RD splitting to any one computed quantity underscores the need for further investigations into this topic. A deeper understanding of this phenomenon would aid the development of spintronic applications of perovskites, particularly for 2DPKs. To this end, we take an initial step towards such investigations with a machine learning model, classifying candidate materials with $\alpha_{RD} > 1.0$ eV Å described in the following section.

## Machine Learning Models for $E_{gap}$, $\alpha_{RD}$, and Atomic Partial Charges

To support the rapid materials exploration, we use the present dataset to develop three machine learning (ML) models geared towards 1) evaluating band gaps to assess the performance of photovoltaic materials, 2) identifying candidate materials for RD splitting, and 3) predicting atomic partial charges to aid in classical simulations of 2DPKs. All performance metrics pertaining to the ML models discussed herein are provided for out-of-sample data, unseen by the model during training. Furthermore, to avoid overfitting, the finalized feature set used for the electronic band gap and RD splitting model was decided upon after analyzing the impact of feature dropout on the out-of-sample test metrics. Features which did not correspond to an improvement on out-of-sample test metrics were removed.

After training and optimizing hyperparameters of various regression models (Supplementary Table S4), we find the GradientBoostingRegressor (GBR) to display the best performance for predicting electronic band gaps in the newly presented 2DPK dataset. Across all six cations and four halogens, we find that electronic band gaps can be determined within a mean absolute error (MAE) of 0.313 eV using only structural and chemical information (Fig. 7a). The full feature set for the $E_{gap}$ and $\alpha_{RD}$ models can be found in Supplementary Table S6. To aid in the physical interpretability of the band gap prediction model, we conduct SHAP analysis.[73] The impact of selected features on the model are presented in the beeswarm plot (Fig. 7b), where colors denote the feature value and placement on the x-axis denotes its impact on the model output. Similar to the previous discussion on RD splitting, the electronegativity difference, $X_{BX}$ Diff, can be seen to have the largest impact on the band gap model. Figure 7b demonstrates that these strongly ionic interactions lead to significant increases in electronic band gaps as predicted by the ML model, consistent with prior analysis on oxides by Di Quarto *et. al*.[74] In other words, maintaining the similar electronegativity at the B- and X-site correlates with both increased RD splitting and lower, more favorable electronic band gaps. The SHAP analysis in Fig. 7b provides additional information, including the observation that lower density leads to lower band gaps. This is also the case with octahedral distortion parameters $\Sigma$, $\Delta$, and $\Lambda_2$, whereby reduced distortions result in lower predicted outputs. This model is useful not only for aiding in physically interpreting key characteristics associated with determining band gaps in 2DPK systems, but also for future studies aiming to target either photovoltaic applications, or those requiring band gaps within a desired range such as LED technologies. With LEDs for example, where narrowband emission at targeted wavelengths is desirable, combining structural information such as octahedral distortions with the predicted band gap serves as a useful tool for generating candidate materials for synthesis.

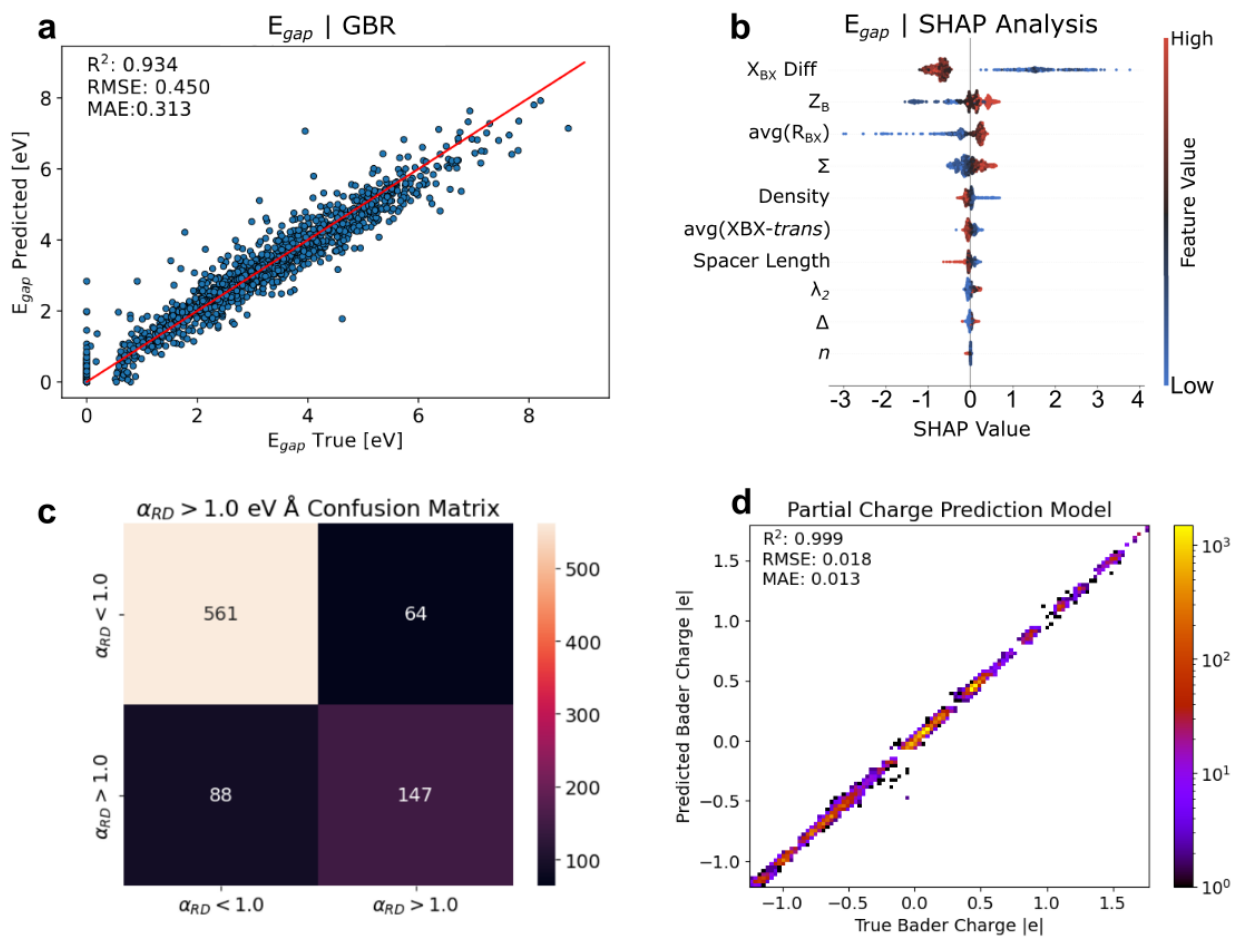

Figure 7. (a) Performance (across cross-validation predictions) of the GradientBoostingRegressor model for prediction of $E_{gap}$ with performance metrics in the inset, (b) SHAP analysis for physical interpretability of the model, (c) confusion matrix for the modified classification model (across cross-validation predictions) predicting candidates with $\alpha_{RD} > 1.0$ eV Å, (d) performance of the HIP-NN model on the held-out test data for partial charge prediction with model metrics in the inset.

To facilitate future investigations into 2DPKs as a platform for the realization of RD splitting, we train a classification model using the results obtained from RashbaPy regarding the $\alpha_{RD}$ splitting of the perovskites in the presented dataset. Figure 7c shows the confusion matrix from the classification model of materials exhibiting $\alpha_{RD} > 1.0$ eV Å. In the same fashion as the $E_{gap}$ regression models, we consider several classification models for which optimized hyperparameters and performance metrics can be found in Supplementary Table S5. The RandomForestClassifier (RF) gives the best performance with an accuracy of 82.4% and an F1-score of 0.66 in classifying the high $\alpha_{RD}$ candidates. To make the model more suitable to identifying candidate materials for subsequent high-throughput searches, we have taken the RF model and modified the bias of the decision tree to mitigate false negatives. This allows the model to miss only 16% of the $\alpha_{RD} > 1.0$ eV Å candidates as opposed to 37% in the unbiased RF classifier. The accuracy of the modified model falls to 81.9%, but the F1-score improves to 0.72, indicating improved performance given the unbalanced dataset. Feature importances (Supplementary Fig. S11) aid in elucidating key features which determine RD splitting in the 2DPK dataset. As previously mentioned, the impact of the B-site cation (and thereby the strength of SOC) is the key feature for $\alpha_{RD}$ predictions, however, the model also captures the impacts of geometric descriptors The standard deviation of octahedral BX bonds $\sigma(R_{BX})$, maximum BX bond length $\max(R_{BX})$, and *trans* X-B-X bond angles are three such examples that also contribute largely to the model predictions. Conclusions from these feature importances are two-fold, 1) significant SOC within the material is a necessary criterion for RD splitting, and 2) octahedral distortions as well as the electronegativity difference between the B- and X-site play the most prominent role in determining whether or not RD splitting will be observed. Of the ten features with highest importance for the $\alpha_{RD}$ model, eight correspond directly to the geometry of the $BX_6$ octahedra present in the 2DPK. It is important to note that these features in experimentally synthesized systems are inherently dynamic, offering an explanation why intrinsic RD splitting is so rarely experimentally observed in materials. These results imply that despite the screening of a large number of 2DPK structures for RD splitting, we have yet to uncover a simple relationship in order to guide the focused discovery of high RD materials in the space of 2DPKs, and that future high-throughput screening approaches are still necessary for the elucidation of such structure property relationships. However, this model and the structural data presented in this dataset can be used to easily impute structural characteristics for a broad range of 2DPKs and drastically reduce the computational cost of workflows aiming to data mine additional 2DPKs exhibiting RD splitting across a broad swath of the periodic table.

The final ML model serves to predict atomic partial charges, using Bader charges computed from the corresponding reference ground state DFT calculations. The computation and prediction of atom-condensed charges (such as Bader charge) is of much use to both classical simulations which employ fixed partial charge, as well as in the generation of reference data for the development of flexible charge models akin to those used in the family of charge equilibration models. This model was trained on all 2104 DFT optimized geometries in the dataset, in contrast with the $E_{gap}$ and $\alpha_{RD}$ models (see Supplemental Information for a full discussion). This model was developed based on the HIP-NN architecture, a deep learning framework which constructs atomic features from the local chemical environment of each atom.[75] The performance on the held-out test data can be seen in Fig. 7d, with MAEs of 1.3% of an electron demonstrating the high accuracy of the model across a diverse set of chemical components spanning 20 elements. Outliers with an error of greater than 0.2 |e| consist entirely of iodide ions which exhibit much closer I-H bond distances than those typical of the dataset. These instances correspond to geometries which exhibit fragmented spacer molecules in favor of the protonation of nearby iodine atoms. With this being the case, these outliers indicate an unstable chemical environment associated with the A'-cation, rather than poor performance by the model in assessing charges associated with chemically feasible geometries. This model may find utility in classical force field simulations, chemical analysis of 2DPK structures obtained from experiment or approximate DFT level methods (such as density functional tight binding methods), and as additional data to improve models predicting characteristics like $\alpha_{RD}$ and $E_{gap}$, both of which can be improved upon the inclusion of atomic partial charges.

## Conclusion

This study presents a dataset of two-dimensional perovskites spanning the Dion Jacobson and Ruddlesden-Popper phases consisting of 2014 compounds with an extensive list of computed properties using density functional theory which is available at https://github.com/r2stanton/2DPK_Database. These include optimized geometries, electronic band gaps and band structures, density of states, the ground state density, frontier orbitals, thermoelectric properties, band structures, and more as detailed in Supplementary Table S2. Our dataset identifies 130 2DPK structures with band gaps within 0.4 eV of the idealized 1.34 eV band gap, making them suitable for photovoltaic applications. Furthermore, the dataset demonstrates the promise of 2DPKs containing Group 14 B-site cations for utilization in mixed photovoltaic-thermoelectric devices. We have also created the RashbaPy package, which to our knowledge is the first code facilitating the high-throughput assessment of Rashba-Dresselhaus splitting in systems which are not limited to 2DPKs (https://github.com/r2stanton/rashbapy). Finally, using the new dataset we trained three machine learning models geared towards 1) aiding future investigations which require 2DPKs with band gaps within some desirable range, 2) facilitating high-throughput studies to identify high $\alpha_{RD}$ materials, and 3) predicting atomic partial charges for computational simulations or the chemical analysis of experimentally obtained structures. This work provides an extensive benchmark dataset of 2DPKs with fully represented spacer molecules, while simultaneously exploring a broad range of chemical compositions within a unified computational framework. The set of optimized crystal structures presented here also forms the basis for future investigations aiming to employ further electronic structure calculations for the deformation potential theory-based analysis of additional thermoelectric and elastic properties in 2DPKs using the approaches we have employed before.[11,76]

## Data availability

The data discussed throughout the manuscript and Supplementary Tables S2, S6 is available at . Real space electron densities and VBM, CBM wavefunctions are available upon request.

## Code availability

The RashbaPy source code is available at https://github.com/r2stanton/rashbapy, and also obtainable through the Python Package Index via the pip command line tool. The repository contains code for generating k-point meshes, quadratically interpolating *ab initio* dispersion relations, automatic detection and computation of $\alpha_{RD}$, and the computation and visualization of spin textures.

## Acknowledgements

R.S. and D.J.T. gratefully acknowledge support from the U.S. National Science Foundation (ECCS-2138728). This work was performed, in part, at the Center for Integrated Nanotechnologies, an Office of Science User Facility operated for the U.S. Department of Energy (DOE) Office of Science by Los Alamos National Laboratory (Contract 89233218CNA000001) and Sandia National Laboratories (Contract DE-NA-0003525). We also acknowledge the Chicoma at LANL and the *Paratpara* at Clarkson University for additional computing resources. LA-UR-24-29922.

## Author contributions

R.S. and D.J.T. formulated the project and computational framework. R.S. performed the computations and wrote the manuscript. W.N. and S. T. provided critical feedback. W.N., S.T., and D.J.T. supervised and directed the work, helped shape the research and analysis and contributed to the final manuscript.

## Competing interests

The authors declare no competing interests.

*Supplementary material for*

# Data Mining and Computational Screening of Rashba-Dresselhaus Splitting and Optoelectronic Properties in Two-Dimensional Perovskite Materials

Robert Stanton[1*], Wanyi Nie[2,3], Sergei Tretiak[3,4,5], and Dhara J. Trivedi[1*]

[1] *Department of Physics, Clarkson University, Potsdam, New York 13699, USA*

[2] *Department of Physics, University at Buffalo, Buffalo, New York 14260, USA*

[3] *Center for Integrated Nanotechnologies, Los Alamos National Laboratory, Los Alamos, NM 87545, USA*

[4] *Theoretical Division, Los Alamos National Laboratory, Los Alamos, NM 87545, USA*

[5] *Center for Nonlinear Studies, Los Alamos National Laboratory, Los Alamos, NM 87545, USA*

AUTHOR INFORMATION

* Corresponding authors: dtrivedi@clarkson.edu and robertmstanton@proton.me

# Computational Methods

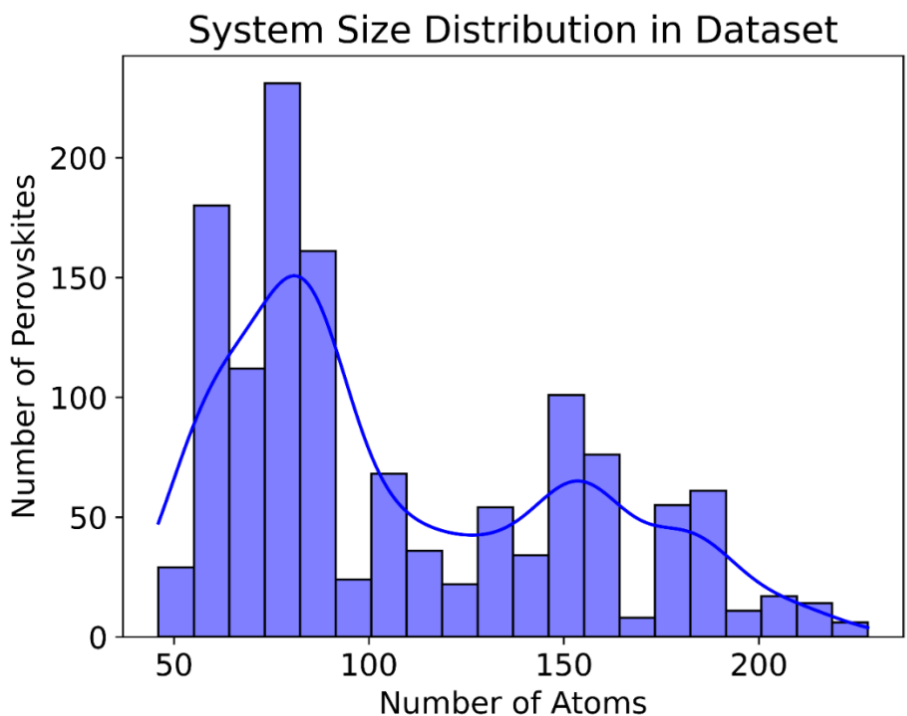

*Figure S1 System size distribution for the 2DPK systems generated in the present study.*

## General Computational Details

We employed a multiscale computational framework to analyze perovskites presented in this work. Initial geometries for all [001] orthorhombic perovskite structures were generated using the freely available pyrovskite code.[1] Subsequently we carry out a variable cell optimization at the xTB level using CP2K.[2] Following a successful geometry optimization at the xTB level, a multi-step geometry optimization at the DFT level with all computational parameters detailed in Figure S2. All DFT calculations used in the present study were done using the GPAW package.[3–5] The BFGS and FIRE optimization algorithms were alternated between for difficult to converge structures, and all 2104 DFT optimized geometries were successfully optimized to within converged forces of .02 eV/Å with a plane-wave kinetic energy cutoff of 500 eV.[6] All structures which failed to converge within this criterion were removed from further consideration. However, we note that the exclusion of these materials does not necessarily entail their structural instability, as some of these systems could likely be converged with more computational resources and fine-tuned electronic structure calculations.

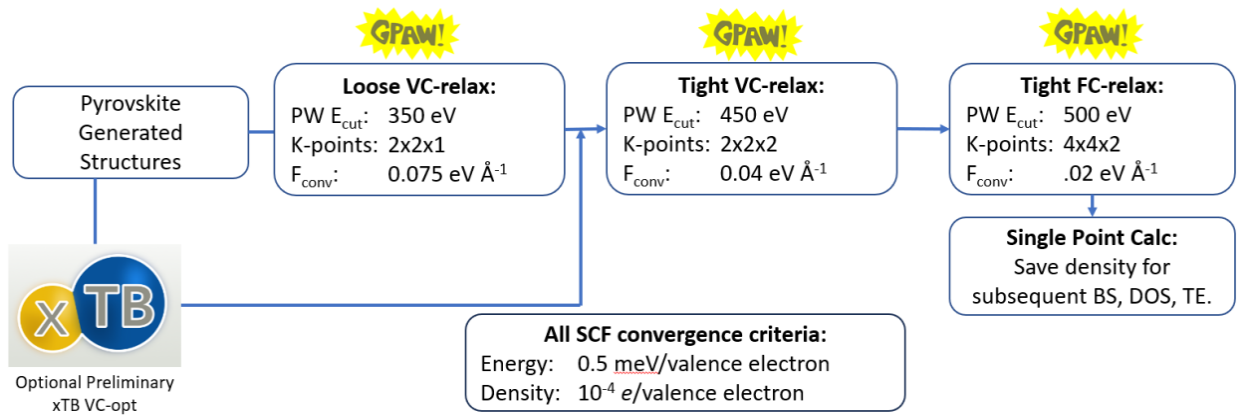

*Figure S2 Computational workflow used for the high-throughput geometry optimization of the 2DPK systems generated in the present study.*

Upon obtaining optimized geometries at the DFT level, one final single point calculation was carried out with an adaptive k-mesh 5 k/Å$^{-1}$, and the ground state density was saved. All subsequent non-self-consistent calculations utilized the ground state density saved in the previous step as an input. Subsequently, we computed the quantities detailed in Table S1. All non-SOC and SOC-included band structure calculations were calculated with an adaptive spacing of 22 k/Å$^{-1}$ along the Setyawan-Curtarolo band path as implemented in the ASE package.[7,8] None of the quantities listed in Table S1 were used as parameters in the machine learning models except those derived from the DFT optimized geometry. One exception is that the Bader atomic partial charges served as the target quantity for the deep learning model for partial charge predictions.[9] An exhaustive list of additional quantities computed and used in the two traditional ML models can be found in Table S5. Further details regarding the computation of RD splitting are discussed in the *Rashba-Dresslehaus Splitting* section.

For the computation of thermoelectric properties, the same mesh used for the non-self-consistent calculation of the density of states is used for the Fourier interpolation of the band energies required by BoltzTraP2. BoltzTraP2 was used with an interpolation multiplier of 3 for all structures in the present study. All Onsager coefficients were computer for carrier concentrations of $\pm 10^{-19} e$/cm$^3$ at 300 K for p- and n-doped configurations respectively.

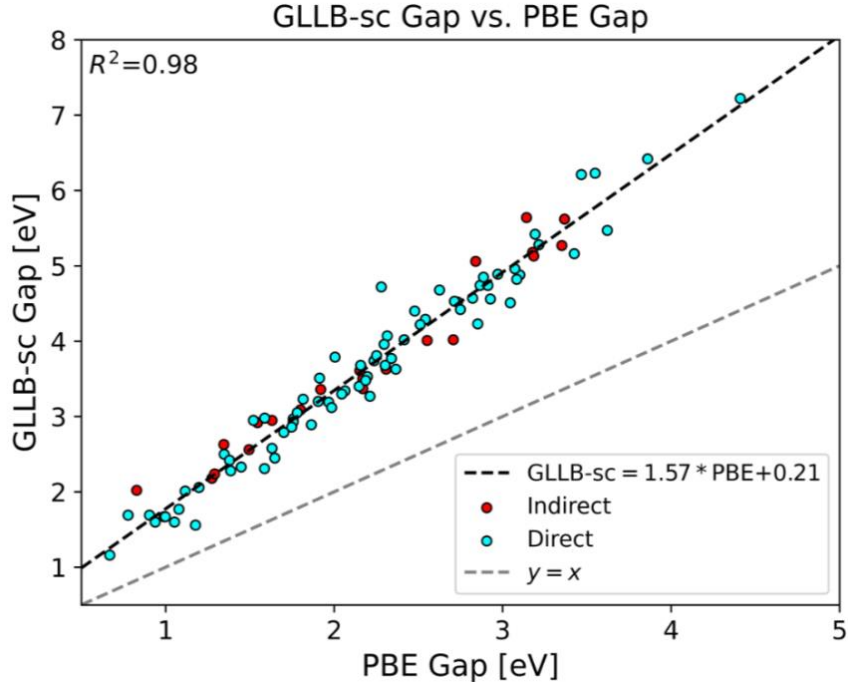

*Figure S3. GLLB-sc band gaps computed for the representative subset in the 2DPK dataset and the resulting linear extrapolation.*

All DFT optimizations were carried out at the PBE level. As discussed in the main manuscript, the computation of band gaps was carried out using the followed equation:

$$E_{gap} = \widetilde{E_{GLLB-sc}} - \Delta_{SOC}$$

In the above equation $\Delta_{SOC}$ is obtained directly as the SOC splitting between the PBE and PBE+SOC calculation as $\Delta_{SOC} = E_{gap}(PBE + SOC) - E_{gap}(PBE)$. $\widetilde{E_{GLLB-sc}}$ is obtained by selecting a subset of the 2DPK dataset comprising a diverse selection of all B-site and X-site ions, as well as a representative sample of directly and indirectly gapped materials. We then utilize the GLLB-sc functional to compute electronic band gaps for this subset.[10] As noted in our previous study, a linear relationship PBE band gaps and GLLB-sc band gaps is obtained and used to extrapolate $\widetilde{E_{GLLB-sc}}$ for all gapped 2DPKs in the present dataset (Figure S3). For the purposes of discussion, we have considered gapped perovskites to be those which exhibit $E_{gap}(PBE) > 0.2$ eV. The GLLB-sc functional serves to estimate the derivative discontinuity missed in a typical GGA-DFT calculation, and strictly increases the band gap as a means of correcting the well-established systematic underestimation problem in DFT. PBE, GLLB-sc, and $E_{gap}$ comparisons with experimentally obtained band gaps in the literature can be found in Table S1.

*Table S1. Comparisons of the PBE band gaps, GLLB-sc band gaps, and $E_{gap} = \widetilde{E_{GLLB-sc}} + \Delta_{SOC}$ with reference experimental band gaps from the literature. The Pb-I and Sn-I rows represent averages of the gapped DJ perovskites of the corresponding layer thickness for comparisons where the specific perovskite from the literature was not available. The nearest approximation to the reference value is underlined. *Comparison is averaged with gapped*

*dataset entries consisting of the same B-cation, halogen, and layer thickness as the reference, while organic A-, A'-cations may differ. A'- and A-cation differences commonly lead to band gap differences up to ~1 eV.*

| Perovskite | n | PBE [eV] | GLLB-sc [eV] | $E_{gap}$ [eV] | Reference |
|---|---|---|---|---|---|
| Pb-I | 1 | 2.08 | 3.47 | <u>2.65</u> | 2.43 [11] |
| Pb-I | 2 | 1.89 | 3.18 | <u>2.18</u> | 2.17 [11] |
| Pb-I | 3 | 1.86 | 3.12 | <u>2.10</u> | 2.03 [11] |
| Pb-I | 4 | 1.52 | 2.59 | <u>1.60</u> | 1.85 [11] |
| Sn-I | 1 | 1.28 | 2.21 | <u>1.95</u> | 1.80 [12] |
| BZA$_2$PbI$_4$ | 1 | 1.80 | 3.03 | 2.10 | 2.18 [13] |
| BZA$_2$SnI$_4$ | 1 | 1.00 | <u>1.78</u> | 1.49 | 1.89 [13] |
| CspFBA$_2$PbI$_4$ | 2 | <u>1.62</u> | 2.76 | 1.76 | 1.61 (n=5) [14] |
| mPDAPbI$_4$ | 1 | 1.87 | 3.15 | <u>2.50</u> | 2.42 [15] |
| mPDAMA Pb$_2$I$_7$ | 2 | 1.91 | 3.20 | <u>2.17</u> | 2.16 [15] |
| mPDAMA Pb$_3$I$_{10}$ | 3 | 1.80 | 3.04 | <u>2.07</u> | 2.00 [15] |
| BA$_2$SnI$_4$ | 1 | 1.21 | 2.11 | <u>1.82</u> | 1.53 [16] |
| BA$_2$SnBr$_4$ | 1 | <u>1.55</u> | 2.64 | 2.42 | 1.98 [16] |
| BA$_2$SnCl$_4$ | 1 | <u>2.06</u> | 3.45 | 3.22 | 2.49 [16] |
| MA$_2$CdCl$_4$* | 1 | 3.07 | <u>5.03</u> | 5.03 | 4.13 [17] |
| MA$_3$Pd$_2$I$_7$* | 2 | 0.39 | <u>0.83</u> | 0.71 | 1.79 [18] |
| MA$_2$PdCl$_4$* | 1 | 1.20 | <u>2.10</u> | 2.03 | 2.15 [19] |
| PEA$_2$GeI$_4$* (compared with n=2) | 1 | 1.42 | 2.44 | <u>2.18</u> | 2.12 [20] |
| PEA$_2$CsPb$_2$Br$_7$* | 2 | <u>2.20</u> | 3.67 | 2.73 | 2.41 [21] |
| Cs$_2$ZnBr$_4$* | 1 | <u>3.54</u> | 5.77 | 5.70 | 4.07 [22] |
| Cs$_2$HgBr$_4$* | 1 | 1.33 | <u>2.29</u> | 2.25 | 3.2-3.3 [23] |

Table S2. Quantities computed for the systems presented in the current 2DPK dataset. Here we do not include quantities which are computable rapidly such as bond distances, distortion parameters, etc. These quantities, many of which were used as features in the machine learning models presented in this study, are available in Table S6.

| Computed Quantity | Identifier/Filename | Comments |
|---|---|---|
| DFT Optimized Geometry | ID.cif | DFT optimized geometry of the perovskite system |
| Density of States | dos_non_soc_ID.json | JSON file containing all DOS information. |
| Band Structure | bs_ID.json | JSON file containing all band structure information. |
| SOC Density of States | soc_dos_ID.json | JSON file containing SOC DOS information. |
| SOC Band Structure | soc_bs_ID.json, ekn_ID.npy, sknv_ID.npy | JSON file containing SOC band structure information. Numpy arrays containing eigenvalues and spin projections. |
| Bader Charges | ACF_ID.dat, AVF_ID.dat, BCF_ID.dat | Bader charge information computed using Bader Charge Analysis from the Henkelman group. |
| VBM WFC | VBM_ID.cube | Cube file from GPAW |
| CBM WFC | CBM_ID.cube | Cube file from GPAW |
| $\rho_{ground-state}$ | density_ID.cube | Cube file from GPAW |
| Onsager Coefficients (Thermoelectric Properties) | interpolation.dope_holes.condtens, interpolation.dope_electrons.condtens | Conductivity tensor information for electron and hole doping. Computed using BoltzTraP2. |

# Structural Properties

Figures S3-S5 display a number of structural characteristics which benefit from further elaboration regarding their computation and their typical values in 2DPK systems. The average *trans* B-X-B angle corresponds to the three X-B-X angles formed in a $BX_6$ octahedra by X-ions opposite one another. In an idealized octahedra of perfect symmetry, these values are 180°, but deviations to around 160 are common in the literature. The *cis* X-B-X angles represent the remaining X-B-X angles which are 90° in idealized octahedra. Average B-X bond lengths naturally vary given the ionic radii of the constituent materials.

The $\Sigma$ octahedral distortion parameter is computed as follows:

$$\Sigma = \frac{1}{6} \sum_i |\phi_i - 90°|$$

Here $\phi_i$ is the *i*-th *cis* X-B-X bond angle in the system. Naturally the $\Sigma$ parameter describes systems which deviate largely from right angles in the *cis* X-B-X bond angles. The $\Delta$ parameter similarly computes a deviation from standard B-X bond lengths within an octahedra as follows:

$$\Delta = \frac{1}{6} \sum_i \left(\frac{d_i - \bar{d}}{d}\right)^2$$

Here $d_i$ is a specific B-X bond distance, and $\bar{d}$ is the average BX bond distance in the octahedra. The $\Lambda_2$ parameter represents dislocations of the B-site cations in a diagonal fashion from the center of their octahedra. A fully detailed derivation and explanation of this parameter can be found in Stanton and Trivedi.[1] All distortion parameters, as well as standard geometric parameters were computed with the Pyrovskite package.

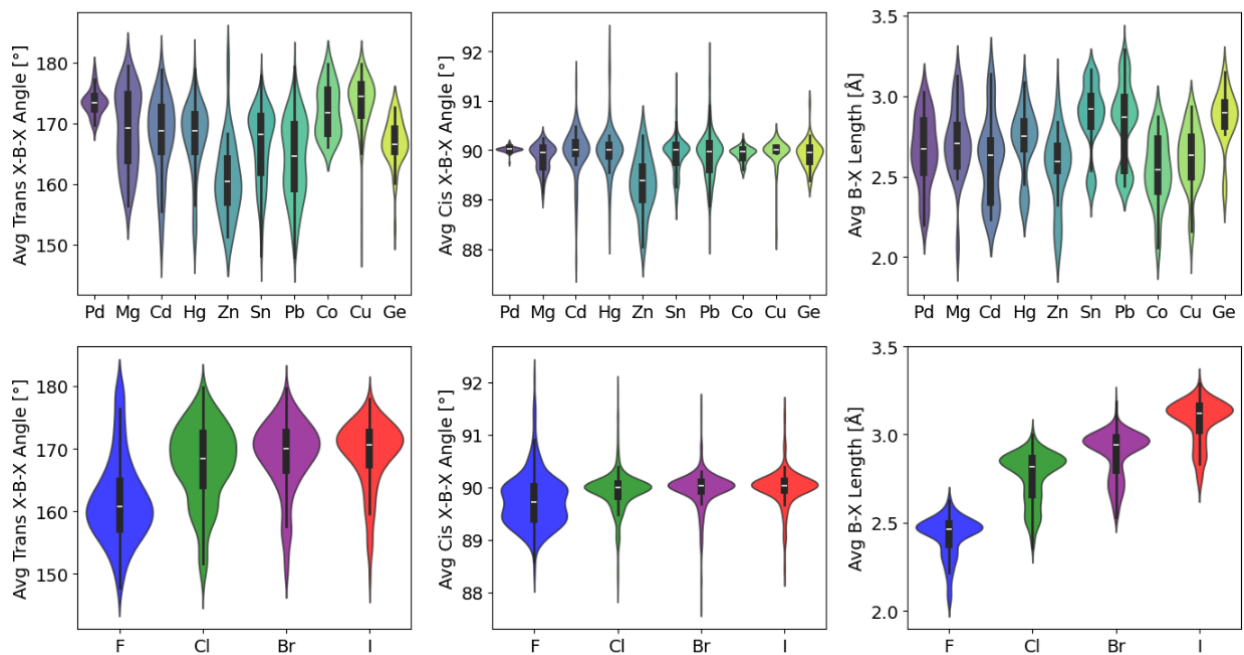

Figure S4. Bond and angle distortions by B-cation and X-halogen for the 2DPK systems generated in the present study.

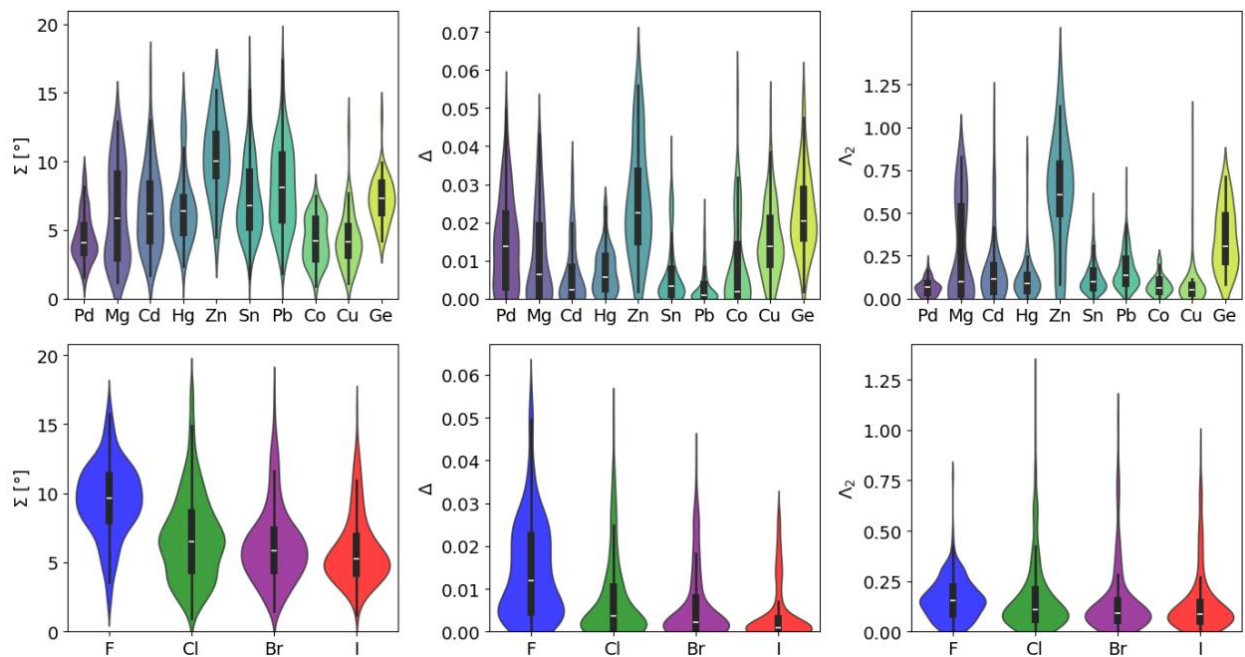

Figure S5 Octahedral distortion parameters for the 2DPK systems generated in the present study.

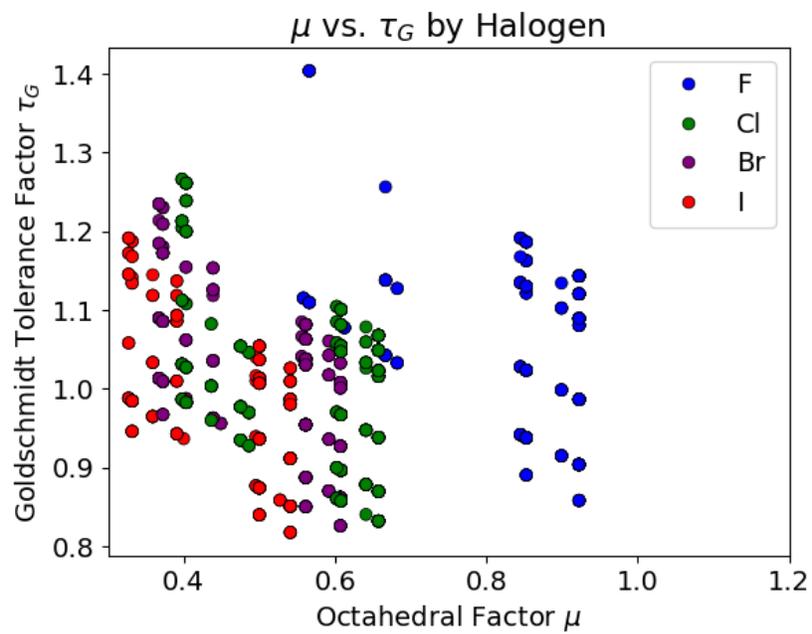

Figure S6 Octahedral factor $\mu$ vs. Goldschmidt tolerance factor $\tau_g$ by halogen for the 2DPK systems generated in the present study.

# **Electronic Properties**

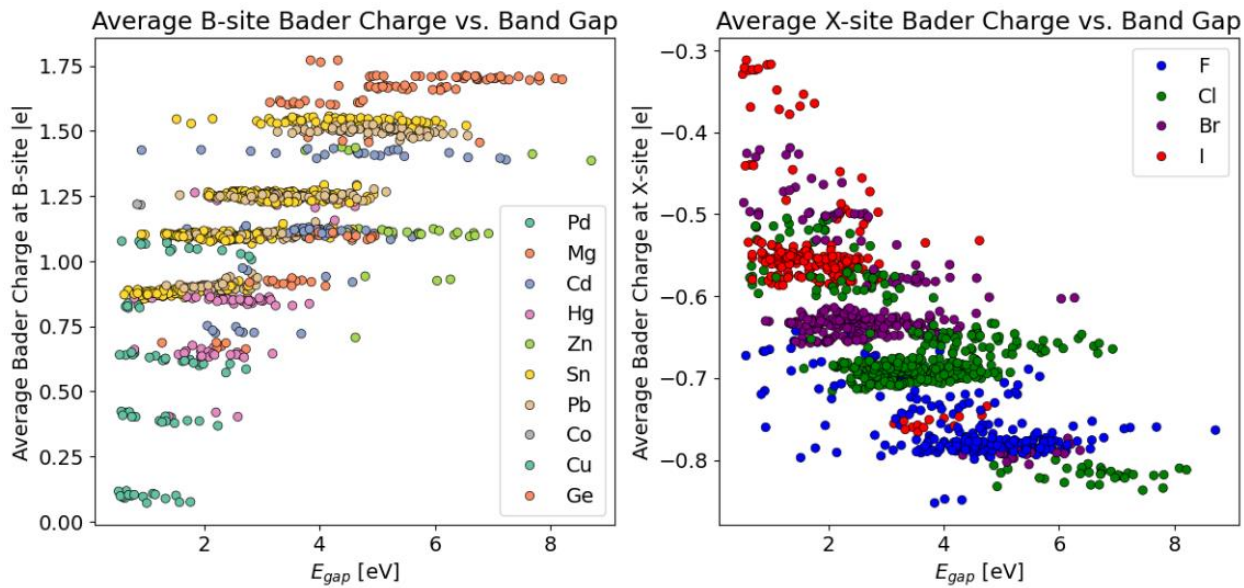

*Figure S7 Electronic band gaps vs. average Bader charges of the B-site and X-site ions for the gapped 2DPKs in the present study.*

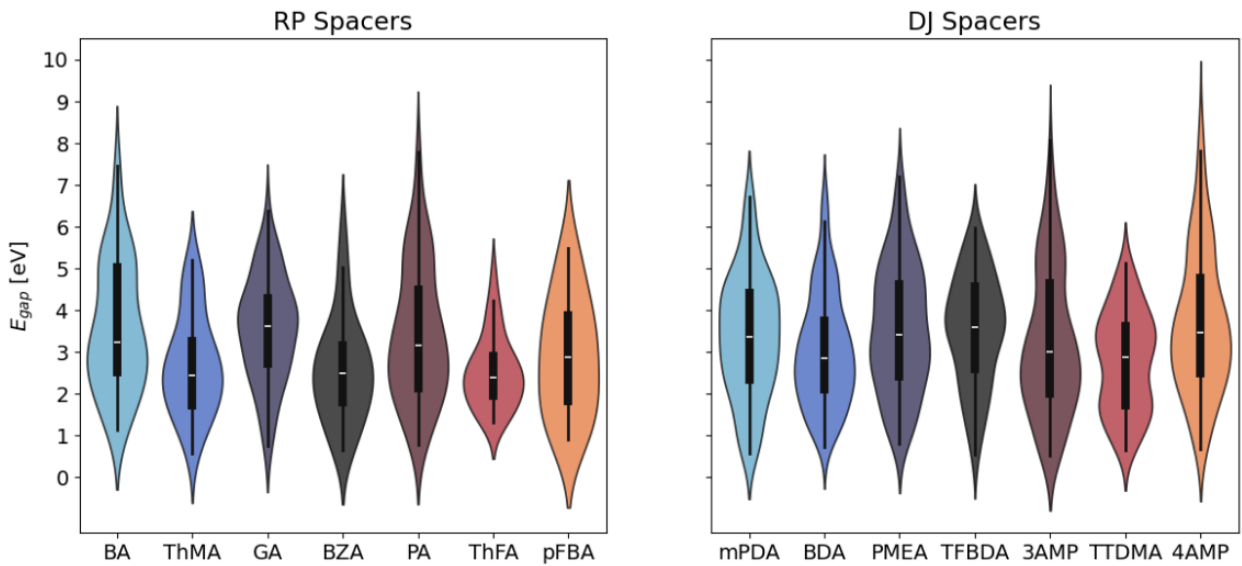

*Figure S8 Band gap distributions for the gapped perovskites in the present dataset for each of the 14 spacers used in this study.*

# **Thermoelectric Properties**

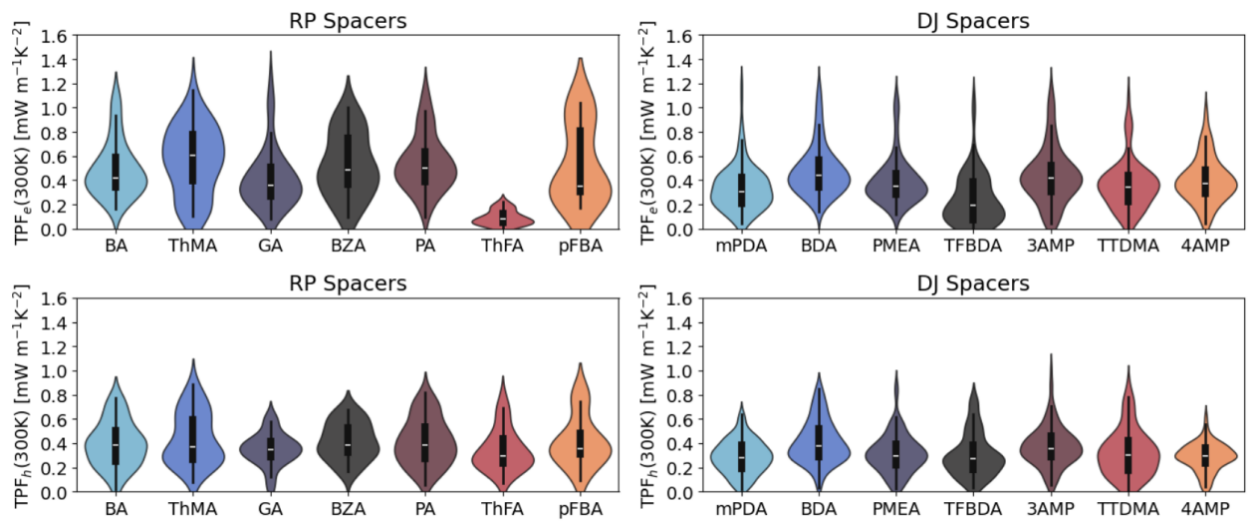

*Figure S9 Thermoelectric power factor associated with n-doped (a, b) and p-doped (c,d) 2DPK systems at carrier concentrations of $\pm 10^{19}$ e/cm$^3$ by spacers in the RP and DJ phases.*

# Rashba-Dresselhaus Splitting

Using the RashbaPy code, we computed Rashba-Dresselhaus splitting ($\alpha_{RD}$) for perovskites with $E_{gap} > 0.5$. The linear combinations of atomic orbitals (LCAO) approach with a triple-$\zeta$ basis set was used within GPAW to ensure that a sufficiently dense *ab initio* k-point sampling of 30 k-points/Å$^{-1}$ could be used for computation of the band structure.[3,4,24] The RashbaPy code carries out a subsequent quadratic interpolation of all *ab initio* data from which the k-point offset ($dk$), and energy offset ($dE$) are computed. All structures with $\alpha_{RD} > 1.0$ eV Å had their crystal structure, non-SOC band structure, interpolated SOC band structure, and spin texture inspected manually. Those systems which exhibited a gapped to gapless transition upon inclusion of SOC were removed from consideration in the statistics presented in the paper as well as the machine learning model. Additionally, systems with indicators of broken octahedra or other distortion that resulted in a non-perovskite like DFT optimized geometry were removed from consideration of both the $\alpha_{RD}$ and $E_{gap}$ models.

An additional challenge in the high-throughput assessment of $\alpha_{RD}$ is attributable to the ongoing discussion in the literature as to what qualifies as true SOC-induced spin-splitting. For example, systems which exhibit band edges offset from a high symmetry point may simply correspond to an indirectly gapped material rather than one displaying RD splitting. However, the literature also contains many examples of systems such as this which are discussed as exhibiting RD splitting. This ambiguity is exacerbated by the fact that, while the spin texture of pure Rashba and Dresselhaus splitting are very easy to identify visually, many 2DPK systems display a hybridization of the two. This greatly complicates distinguishing candidate RD splitting materials from "false-positives" through visual inspection of the spin texture. For all 2DPKs with $\alpha_{RD} > 1$, any systems which have band edges offset from high symmetry points that did not appear to be exacerbated by the inclusion of SOC were removed from further consideration. This serves the purpose of reducing falsely identifying standard indirectly gapped materials as $\alpha_{RD}$ splitting candidates. With this being the case, we acknowledge that upon more fine-grained investigations of the high $\alpha_{RD}$ materials presented in this study, it could be the case that some of these do not truly exhibit RD splitting. No further consideration was given to the delineation between Rashba and Dresselhaus splitting.

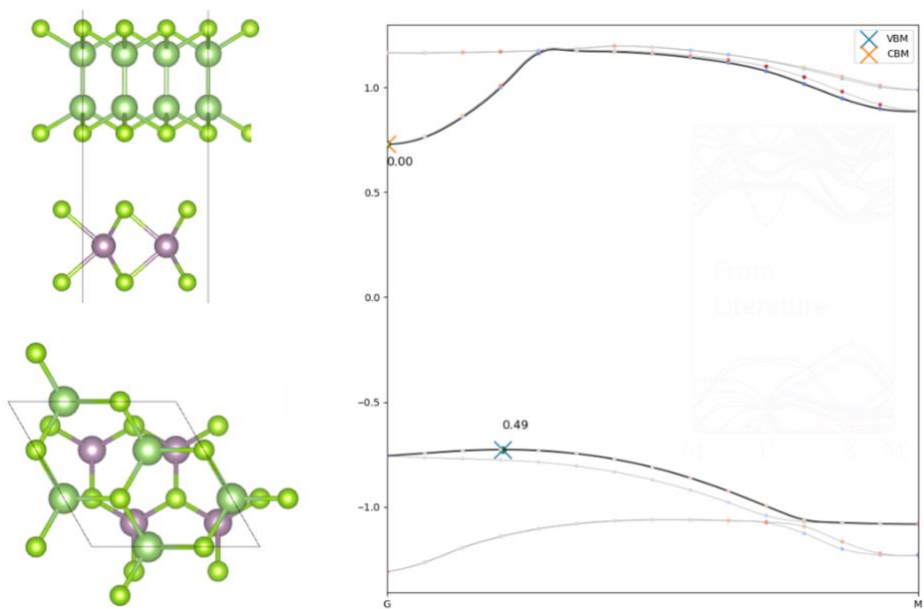

*Figure S10. Benchmark of Rashba-Dresselhaus splitting in the GaSe/MoSe$_2$ system from Zhang and Schwingenschlögl. Our high-throughput friendly approach comes to an identical $\alpha_{RD}$ parameter of 0.49 eV Å in the VB and 0.0 eV Å in the CB.*

*Table S3. Reference values used to benchmark the RashbaPy code used for the high-throughput computation of R-D splitting in all of the selected 2DPK systems.*

| Benchmarked Quantity | Our Value | Reference Value[25] |
| --- | --- | --- |
| $\alpha_{RD}$ [eV Å] | 0.49 | 0.49 |
| dE [meV] | 31 | 30 |
| dk [Å$^{-1}$] | 0.13 | 0.12 |

# Machine Learning Models

All machine learning models for the prediction of electronic band gaps and Rashba-Dresselhaus splitting were trained using sci-kit learn.[26] Table S4 and Table S5 contain the optimized hyperparameters GridSearchCV, as well as the metrics for the best performing model, and final optimized parameters. All performance metrics are computed from 20-fold cross-validation predictions. Table S6 contains all features used in the two models as well as a brief description of the feature. All features are scaled with the StandardScaler from sci-kit learn, shifting the mean to 0 and variance to 1. All features which were not computed manually or with the pyrovskite package were done using the Matminer package.[27]

The HIP-NN architecture as implemented in the hippynn package was used for the deep learning model for partial atomic charge prediction with Bader charges used as reference data from the ground state DFT calculations described above.[28] Network parameters consisted of two interaction layers, separated by three on-site layers, and 54 features per input atom. The loss function used in the training was the sum of the mean absolute error between predicted charges, the root mean square error of charge predictions, and an L2 regularization term to prevent overfitting.

$$\mathcal{L} = MAE_q + RMSE_q + L_2$$

Performance metrics for the HIP-NN model are provided in Table S7. While the $E_{gap}$ and $\alpha_{RD}$ models can be used for 2DPKs constituting any element (though likely with decreased accuracy for those outside of the dataset), the HIP-NN model can only be used for those within the dataset. This restricts the usage of the HIP-NN model to systems containing H, C, N, F, Mg, S, Cl, Co, Cu, Zn, Ge, Br, Rb, Pd, Cd, Sn, I, Cs, Hg, and Pb. Though we note that this is a relatively extensive list of elements that covers the overwhelming majority of 2DPKs found in the literature, with a notable exception regarding organic spacers containing O-atoms, such as amin-acids frequently employed in chiral perovskites.

*Table S4. Table containing the optimized hyper parameters for all models trained for predicting electron band gaps. All evaluation metrics are from 20-fold cross-validation predictions on unseen data by the trained models. The GradientBoostingRegressor was used for all SHAP values, owing to ease of interpretability and only marginally worse performance than the VotingRegressor. *The VotingRegressor comprises the ExtraTrees, GradientBoosting, RandomForest, and MLP regressors trained on their respective optimized hyperparameters.*

| Model Type | Optimized Hyperparameters | $R^2$ | RMSE [eV] | MAE [eV] |
|---|---|---|---|---|
| RandomForestRegressor | max_depth: None | 0.913 | 0.516 | 0.352 |
|  | n_estimators: 225 |  |  |  |
|  | max_features: 0.9 |  |  |  |
| GradientBoostingRegressor | learning_rate: 0.1 | 0.934 | 0.450 | 0.313 |
|  | max_depth: 5 |  |  |  |
|  | n_estimators: 200 |  |  |  |
| BaggingRegressor | max_samples: 0.9 | 0.908 | 0.530 | 0.365 |
|  | n_estimators: 30 |  |  |  |

| Model Type | Optimized Hyperparameters | | | |
|---|---|---|---|---|
| ExtraTreesRegressor | max_depth: None | 0.920 | 0.493 | 0.332 |
| | n_estimators: 100 | | | |
| MLP | hidden_layer_sizes: [16, 16, 16] | 0.907 | 0.532 | 0.369 |
| | alpha: 0.001 | | | |
| | batch_size: 16 | | | |
| | learning_rate_init: .01 | | | |
| | learning_rate: 'constant' | | | |

*Table S5. Table containing the optimized hyperparameters for all models trained for predicting candidate materials with $\alpha_{RD} > 1.0$ eV Å. A VotingClassifier was not included here because it does not offer improved predictions over the best performing individual model. *The RandomForestClassifier+Bias entry represents the same parameters as the RandomForestClassifier with a modified cutoff to mitigate False Negatives in the search for high $\alpha_{RD}$ materials.*

| Model Type | Optimized Hyperparameters | Accuracy | F1-Score |
|---|---|---|---|
| RandomForestClassifier | max_depth: None | 82.4 | 0.66 |
| | n_estimators: 200 | | |
| | max_features: 0.9 | | |
| GradientBoostingClassifier | learning_rate: 0.08 | 81.0 | 0.64 |
| | max_depth: 8 | | |
| | n_estimators: 250 | | |
| BaggingClassifier | max_samples: 0.75 | 80.7 | 0.62 |
| | n_estimators: 30 | | |
| ExtraTreesClassifier | max_depth: None | 82.3 | 0.66 |
| | n_estimators: 25 | | |
| MLP | hidden_layer_sizes: [32, 32, 32] | 81.0 | 0.65 |
| | alpha: 0.001 | | |
| | batch_size: 16 | | |
| | learning_rate_init: .01 | | |
| | learning_rate: 'constant' | | |
| *RandomForestClassifier+Bias | - | 81.9 | 0.72 |

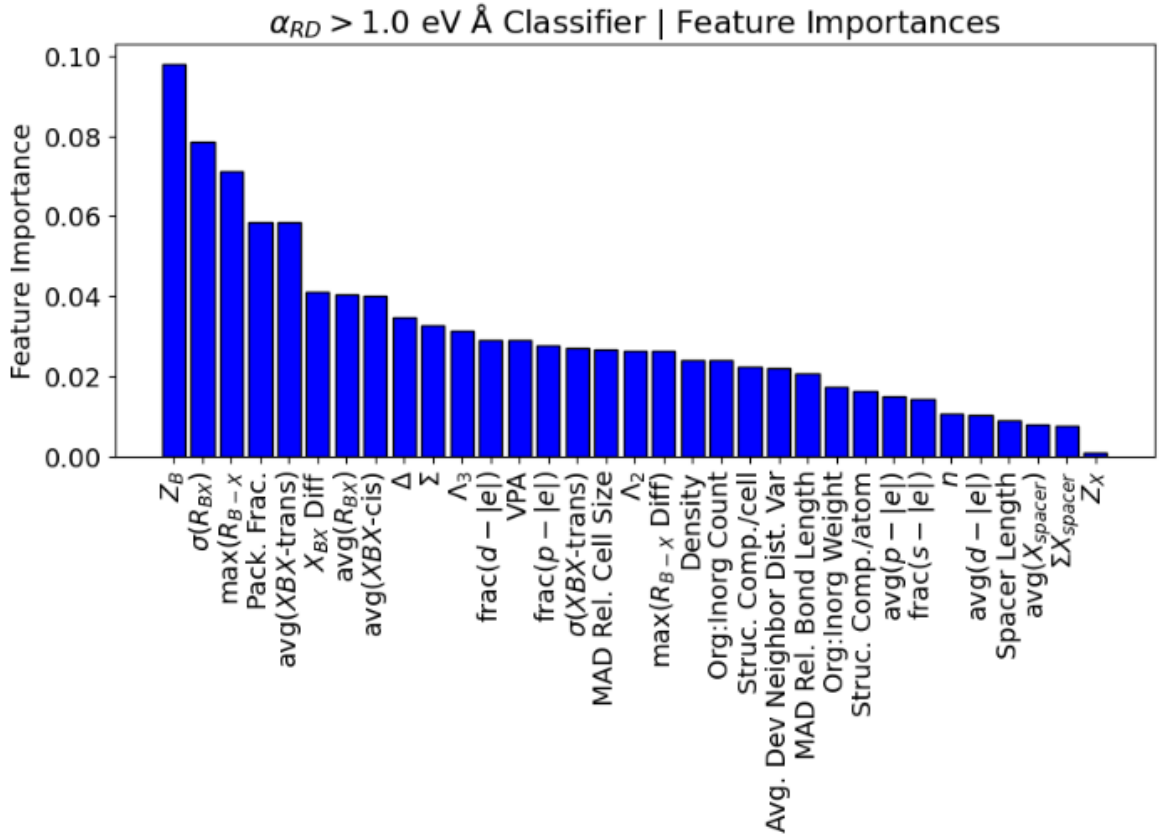

*Figure S11. Feature importances for the classification model aimed at determining candidate material exhibiting $\alpha_{RD} > 1.0$ eV Å.*

*Table S6. All machine learning features used in the training of the model for prediction $E_{gap}$ and $\alpha_{RD}$. These features are computed from a combination of the Pyrovskite and Matminer.*

| ML Feature | Identifier | Description |
| --- | --- | --- |
| $n$ | "n" | Inorganic layer thickness |
| $\Sigma$ | "sigma" | Octahedral distortion parameter (Bond angles) |
| $\Delta$ | "delta" | Octahedral distortion parameter (Bond lengths) |
| $\Lambda_3$ | "lambda3" | Octahedral distortion parameter (B-cation displacement) |
| $\Lambda_2$ | "lambda2" | Octahedral distortion parameter (B-cation displacement) |
| $\max(R_{BX}\text{ Diff})$ | "max_BX_dist_diff" | Maximum distance between largest and smallest BX bond length |
| $X_{BX}$ Diff | "X_BX_diff" | Difference in Pauling electronegativities between X, B components |
| Spacer Length | "spacer_length" | Distance between furthest atoms in spacer |
| Spacer Electronegativity | "spacer_tot_eneg" | Summed Pauling electronegativities of the spacer |

| Avg(Spacer Electronegativity) | "spacer_ave_eneg" | Averaged Pauling electronegativities of the spacer |
|---|---|---|
| max($R_{BX}$) | "max_BX_dist" | Maximum BX bond length |
| Avg(BXB-trans) | "BXB_trans_avg" | Average trans BXB angle in octahedra |
| Avg(BXB-cis) | "BXB_cis_avg" | Average cis BXB angle in octahedra |
| Avg($R_{BX}$) | "BX_avg | Average BX bond length |
| $\sigma(R_{BX})$ | "BX_std" | Standard deviation of BX bond lengths |
| $\sigma$(BXB-trans) | "BXB_trans_std" | Standard deviation of trans BXB angles |
| Struc. Comp./atom | "structural complexity per atom" | Sannon Entropy per atom |
| Struc. Comp./cell | "structural complexity per cell" | Sannon Entropy per cell |
| Density | "density" | Mass density of the perovskite system |
| Volume per atom | "vpa" | Volume per atom |
| Packing fraction | "Pack. Frac." | Packing fraction for the perovskite system |
| avg($p-|e|$) | "avg p valence electrons" | Average number of p-valence electrons |
| avg($d-|e|$) | "avg d valence electrons" | Average number of d-valence electrons |
| frac($s-|e|$) | "avg s valence electrons" | Average number of s-valence electrons |
| frac($p-|e|$) | "avg p valence electrons" | Average number of p-valence electrons |
| frac($d-|e|$) | "avg d valence electrons" | Average number of d-valence electrons |
| MAD Rel. Bond Length | "mean absolute deviation in relative bond length" | MAD in relative bond lengths of the perovskite structure |
| Avg. Dev Neighbor Dist. Var | "avg_dev neighbor distance variation" | Average deviation in neighbor distances |
| MAD Rel. Cell Size | "mean absolute deviation in relative cell size" | MAD relative deviation in relative cell size |
| Org:Inorg Weight | "org_inorg_weight_ratio" | Summed relative weight of organic to inorganic perovskite components |
| Org:Inorg Count | "org_inorg_ratio" | Summed relative count of organic and inorganic atoms in the perovskite system |
| OHE Phase | Phase | OHE for DJ (0), RP (1) phases. |
| Bz | "Bz" | Atomic number for the B-cation |
| Xz | "Xz" | Atomic number for the X-halogen |

*Table S7. Performance metrics for the HIPN-NN model used for partial charge prediction across the full dataset of 2104 optimized DFT geometries.*

| Dataset Split | MAE |e| | RMSE |e| |
|---|---|---|
| Train (80%) | 0.01265 | 0.017871 |
| Validation (10%) | 0.012608 | 0.017212 |
| Test (10%) | 0.012552 | 0.017589 |